%%%%%%%%%%%%%%%%%%%%%%%%%%%%%%%%%%%%%%%%%%%%%%%%%%%%%%%%%%%%%%%%%
\input harvmac
% \draftmode
\noblackbox
%%%%%%%%%%%%%%%%% Lineskip %%%%%%%%%%%%%%%%%%%%%%%%%%%%%%%%%%%%%%%
\ifx\answ\bigans
\magnification=1200\baselineskip=14pt plus 2pt minus 1pt
\else\baselineskip=16pt % plus 2pt minus 1pt % 32 lines in l-format
\fi
%%%%%%%%%%%% Local definitions %%%%%%%%%%%%%%%%%%%%%%%%%%%%%%%%%%%%
\def\vz{{ \{\overline z_n \} }}\def\vzw{{ 
\{z_n \} }}
\def\va{{ \{\overline a_m \} }}\def\vaw{{ \{a_m \} }}

\def\th{\theta}
\def\al{\alpha}
\def\val{{\vec\alpha}} \def\vbe{{\vec\beta}} \def\vde{{\vec\delta}}

\def\bet{\beta}\def\be{\beta}
\def\Om{\Omega}
\def\om#1{\omega_{#1}}
\def\om#1#2{\omega_{#1}(#2)}
\def\zh{{z \over 2}}
\def\hiii{\hskip3mm}
\def\hv{\hskip20mm}

\def\bm{\h+\beta_2}\def\bn{\h+\beta_1}
%%%%%%%%%%%%%%%%%%%%%%%%%%%%%%%%%%%%%%%%%%%%%%%%%%%%%%%%%%%%%%%%%%
%%%%% Referencing
%%%%%%%%%%%%%%%%%%%%%%%%%%%%%%%%%%%%%%%%%%%%%%%%%%%%%%%%%%%%%%%%%%
\newif\ifnref
\def\rrr#1#2{\relax\ifnref\nref#1{#2}\else\ref#1{#2}\fi}
\def\ldf#1#2{\begingroup\obeylines
\gdef#1{\rrr{#1}{#2}}\endgroup\unskip}

\nreffalse

\def\lref{\ldf}
%%%%%%%%%%%%%%%%%%%%%%%%%%%%%%%%%%%%%%%%%%%%%%%%%%%%%%%%%%%%%%%%%%%%%%
%%%%%%%%%%%%%%%   Standard alltime definitions   %%%%%%%%%%%%%%%%%%%%%
%%%%%%%%%%%%%%%%%%%%%%%%%%%%%%%%%%%%%%%%%%%%%%%%%%%%%%%%%%%%%%%%%%%%%%

\def\appA{A}
\def\appAiv{A.4}\def\appAv{A.5}
\def\appB{B}

\def\tilde{\widetilde}
\def\h {{1\over 2}}
\def\ov {\overline}
\def\o {\over}
\def\fc#1#2{{#1 \o #2}}
\def\IZ{ {\bf Z}}
\def\IC{{\bf C}}

\def\hat{\widehat}
\def\E {\hat E}      % For Eisenstein E2
  % For Polylogarithm

\def\br{\hfill\break}
\def\tr {{\rm tr}}
\def\det {{\rm det}}

\def\lf {\left}
\def\ri {\right}
\def\ra {\rightarrow}

\def\re {{\rm Re}}
\def\im {{\rm Im}}
\def\p {\partial}

\def\Fc {{\cal F}} 
 \def\Oc {{\cal O}}

\def\Rc {{\cal R}}
\def\Ic {{\cal I}}
\def\Kc {{\cal K}}
%%%%%%%%%%%%%%%%%%%%%%%%%%%%%%%%%%%%%%%%%%%%%%%%%%%%%%%%%%%%%%%%%%%

%\input refs
\lref\STi{
S.~Stieberger and T.R.~Taylor,
``Non-Abelian Born-Infeld action and type I - heterotic duality  (I): 
Heterotic $F^6$ 
terms at two loops,''
arXiv:hep-th/0207026, to appear in Nucl.\ Phys.\ B.
%%CITATION = HEP-TH 0207026;%%
}

\lref\STiii{
S.~Stieberger and T.R.~Taylor,
``Non-Abelian Born-Infeld action and type I - heterotic duality  (III),'' to 
appear}

\lref\tseytlin{A.A.~Tseytlin,
'`On non-abelian generalisation of the Born-Infeld action in string  theory,''
Nucl.\ Phys.\ B {\bf 501}, 41 (1997)
[arXiv:hep-th/9701125];
%%CITATION = HEP-TH 9701125;%%
``Born-Infeld action, supersymmetry and string theory,''
arXiv:hep-th/9908105.
%%CITATION = HEP-TH 9908105;%%
}

\lref\allAbelian{A.A.~Tseytlin,
``Vector Field Effective Action In The Open Superstring Theory,''
Nucl.\ Phys.\ B {\bf 276}, 391 (1986)
[Erratum-ibid.\ B {\bf 291}, 876 (1987)].}

\lref\sevrin{A.~Sevrin, J.~Troost and W.~Troost,
``The non-abelian Born-Infeld action at order $F^6$'',
Nucl.\ Phys.\ B {\bf 603}, 389 (2001)
[arXiv:hep-th/0101192].
%%CITATION = HEP-TH 0101192;%%
}

\lref\GSi{M.B.~Green and J.H.~Schwarz,
``Supersymmetrical Dual String Theory. 2. Vertices And Trees,''
Nucl.\ Phys.\ B {\bf 198}, 252 (1982).
%%CITATION = NUPHA,B198,252;%%
}

\lref\GSii{M.B.~Green and J.H.~Schwarz,
``Supersymmetrical Dual String Theory. 3. Loops And Renormalization,''
Nucl.\ Phys.\ B {\bf 198}, 441 (1982).
%%CITATION = NUPHA,B198,441;%%
}

\lref\witten{
E.~Witten,
``String theory dynamics in various dimensions,''
Nucl.\ Phys.\ B {\bf 443} (1995) 85
[arXiv:hep-th/9503124];\br
J.~Polchinski and E.~Witten,
``Evidence for Heterotic - Type I String Duality,''
Nucl.\ Phys.\ B {\bf 460} (1996) 525
[arXiv:hep-th/9510169].
%%CITATION = HEP-TH 9510169;%%
}
\lref\review{For a review, see e.g.:
I.~Antoniadis, H.~Partouche and T.~R.~Tay\nolinebreak lor,
``Lectures on heterotic-type I duality,''
Nucl.\ Phys.\ Proc.\ Suppl.\  {\bf 61A} (1998) 58
[arXiv:hep-th/9706211];\br
E.~Kiritsis,
``Introduction to non-perturbative string theory,''
arXiv:hep-th/9708130;\br
%%CITATION = HEP-TH 9708130;%%
A.~Sen,
``An introduction to non-perturbative string theory,''
arXiv:hep-th/9802051.
%%CITATION = HEP-TH 9802051;%%
}

\lref\AA{A.A. Tseytlin,
``On SO(32) heterotic - type I superstring duality in ten dimensions,''
Phys.\ Lett.\ B {\bf 367} (1996) 84
[arXiv:hep-th/9510173];
% A.~A.~Tseytlin,
``Heterotic - type I superstring duality and low-energy effective actions,''
Nucl.\ Phys.\ B {\bf 467} (1996) 383
[arXiv:hep-th/9512081].
%%CITATION = HEP-TH 9512081;%%
}

\lref\BK{
C.~Bachas and E.~Kiritsis,
``$F^4$ terms in N = 4 string vacua,''
Nucl.\ Phys.\ Proc.\ Suppl.\  {\bf 55B}, 194 (1997)
[arXiv:hep-th/9611205].
%%CITATION = HEP-TH 9611205;%%
}

\lref\AGNT{
I. Antoniadis, E. Gava, K.S. Narain and T.R. Taylor,
``Topological amplitudes in string theory,''
Nucl.\ Phys.\ B {\bf 413}, 162 (1994)
[hep-th/9307158];
%%CITATION = HEP-TH 9307158;%%
% I.~Antoniadis, E.~Gava, K.S.~Narain and T.R.~Taylor,
``Topological Amplitudes in Heterotic Superstring Theory,''
Nucl.\ Phys.\ B {\bf 476}, 133 (1996)
[hep-th/9604077].
%%CITATION = HEP-TH 9604077;%%
}

\lref\LSW{W.~Lerche, B.E.~Nilsson, A.N.~Schellekens and N.P.~Warner,
``Anomaly Cancelling Terms From The Elliptic Genus,''
Nucl.\ Phys.\ B {\bf 299}, 91 (1988)\br
%%CITATION = NUPHA,B299,91;%%
W.~Lerche, B.E.~Nilsson and A.N.~Schellekens,
``Heterotic String Loop Calculation Of The Anomaly Cancelling Term,''
Nucl.\ Phys.\ B {\bf 289}, 609 (1987);\br
%%CITATION = NUPHA,B289,609;%%
W.~Lerche,
``Elliptic Index And Superstring Effective Actions,''
Nucl.\ Phys.\ B {\bf 308}, 102 (1988).
%%CITATION = NUPHA,B308,102;%%
}

\lref\LSWgenus{A.N. Schellekens and N.P.~Warner,
``Anomalies And Modular Invariance In String Theory,''
Phys.\ Lett.\ B {\bf 177}, 317 (1986);
%%CITATION = PHLTA,B177,317;%%
``Anomalies, Characters And Strings,''
Nucl.\ Phys.\ B {\bf 287}, 317 (1987);\br
%%CITATION = NUPHA,B287,317;%%
W.~Lerche, A.N.~Schellekens and N.P.~Warner,
``Lattices And Strings,''
Phys.\ Rept.\  {\bf 177}, 1 (1989).
%%CITATION = PRPLC,177,1;%%
}

\lref\yasuda{O.~Yasuda,
``Absence Of Hexagon Gauge Anomalies In Higher Loop Superstring Amplitudes,''
Phys.\ Lett.\ B {\bf 215}, 306 (1988);
%%CITATION = PHLTA,B215,306;%%
``Nonrenormalization Theorem For The Green-Schwarz Counterterm And The 
Low-Energy 
Effective Action,''
Phys.\ Lett.\ B {\bf 218}, 455 (1989);\br
%%CITATION = PHLTA,B218,455;%%}
L.~Mizrachi,
``A Multiloop Analysis Of The Anomaly Cancelling Terms In The Heterotic 
String,''
Nucl.\ Phys.\ B {\bf 338}, 209 (1990).
%%CITATION = NUPHA,B338,209;%%
}

\lref\BPT{I.L. Buchbinder, A.Y. Petrov and A.A. Tseytlin,
``Two-loop N=4 super Yang Mills effective action and interaction  between 
D3-branes,''
Nucl.\ Phys.\ B {\bf 621}, 179 (2002)
[arXiv:hep-th/0110173].
%%CITATION = HEP-TH 0110173;%%
}

\lref\iengo{E.~Gava, R.~Jengo and G.~Sotkov,
``Modular Invariance And The Two Loop Vanishing Of The Cosmological Constant,''
Phys.\ Lett.\ B {\bf 207}, 283 (1988);\br
%%CITATION = PHLTA,B207,283;%%
A.~Morozov,
``On Two Loop Contribution To Four Point Function For Superstring,''
Phys.\ Lett.\ B {\bf 209}, 473 (1988)
[JETP Lett.\  {\bf 47}, 219 (1988)];\br
%%CITATION = PHLTA,B209,473;%%
R.~Jengo, G.M.~Sotkov and C.J.~Zhu,
``Two Loop Vacuum Amplitude In Four-Dimensional Heterotic String Models,''
Phys.\ Lett.\ B {\bf 211}, 425 (1988);\br
%%CITATION = PHLTA,B211,425;%%
R.~Jengo and C.J.~Zhu,
``Notes On Nonrenormalization Theorem In Superstring Theories,''
Phys.\ Lett.\ B {\bf 212}, 309 (1988);
%%CITATION = PHLTA,B212,309;%%
``Two Loop Computation Of The Four Particle Amplitude In Heterotic String 
Theory,''
Phys.\ Lett.\ B {\bf 212}, 313 (1988).
%%CITATION = PHLTA,B212,313;%%
}

\lref\JZ{
R.~Iengo and C.~J.~Zhu,
``Explicit modular invariant two-loop superstring amplitude relevant for  
$R^4$,''
JHEP {\bf 9906}, 011 (1999)
[arXiv:hep-th/9905050].
%%CITATION = HEP-TH 9905050;%%
}

\lref\morozov{A.~Morozov,
``Two Loop Statsum Of Superstring,''
Nucl.\ Phys.\ B {\bf 303}, 343 (1988).
%%CITATION = NUPHA,B303,343;%%
}

\lref\zhu{C.~J.~Zhu,
``Two Loop Computations In Superstring Theories,''
Int.\ J.\ Mod.\ Phys.\ A {\bf 4}, 3877 (1989).
%%CITATION = IMPAE,A4,3877;%%
}

\lref\mumford{D. Mumford, ``Tata Lectures on Theta'',
vols I and II, Birkhauser, Boston 1983.}

\lref\ellisone{
J.R.~Ellis, P.~Jetzer and L.~Mizrachi,
``One Loop String Corrections To The Effective Field Theory,''
Nucl.\ Phys.\ B {\bf 303}, 1 (1988).
%%CITATION = NUPHA,B303,1;%%
}

\lref\clavelli{
L.~Clavelli,
``Proof Of One Loop Finiteness Of Type I SO(32) Superstring Theory,''
Phys.\ Rev.\ D {\bf 33}, 1098 (1986);\br
%%CITATION = PHRVA,D33,1098;%%
L.~Clavelli, P.H.~Cox and B.~Harms,
``Parity Violation In Type I Superstring Graviton Exchange,''
Phys.\ Rev.\ D {\bf 35}, 1385 (1987);
%%CITATION = PHRVA,D35,1385;%%
``Parity Violating One Loop Six Point Function In Type I Superstring Theory,''
Phys.\ Rev.\ D {\bf 35}, 1908 (1987).
%%CITATION = PHRVA,D35,1908;%%
}

\lref\PINCH{M.B.~Green and N.~Seiberg,
``Contact Interactions In Superstring Theory,''
Nucl.\ Phys.\ B {\bf 299}, 559 (1988);\br
%%CITATION = NUPHA,B299,559;%%
J.A.~Minahan,
``One Loop Amplitudes On Orbifolds And The Renormalization Of Coupling 
Constants,''
Nucl.\ Phys.\ B {\bf 298}, 36 (1988);\br
%%CITATION = NUPHA,B298,36;%%
P.~Mayr and S.~Stieberger,
``Dilaton, antisymmetric tensor and gauge fields in string effective theories 
at the 
one loop level,''
Nucl.\ Phys.\ B {\bf 412}, 502 (1994)
[arXiv:hep-th/9304055].
%%CITATION = HEP-TH 9304055;%%.
}

\lref\gross{
D.~J.~Gross and J.~H.~Sloan,
``The Quartic Effective Action For The Heterotic String,''
Nucl.\ Phys.\ B {\bf 291}, 41 (1987).
%%CITATION = NUPHA,B291,41;%%
}

\lref\LP{
O.~Lechtenfeld and A.~Parkes,
``On Covariant Multiloop Superstring Amplitudes,''
Nucl.\ Phys.\ B {\bf 332}, 39 (1990).}
%%CITATION = NUPHA,B332,39;%%

\lref\HP{E. D'Hoker and D.H. Phong,
``Two-loop superstrings. I--IV'',
Phys.\ Lett.\ B {\bf 529}, 241 (2002), 
[arXiv:hep-th/0110247];
%%CITATION = HEP-TH 0110247;%%
% ``Two-loop superstrings. II: The chiral measure on moduli space,''
arXiv:hep-th/0110283;
%%CITATION = HEP-TH 0110283;%%
% ``Two-loop superstrings. III: Slice independence and absence of  
%ambiguities,''
arXiv:hep-th/0111016;
%%CITATION = HEP-TH 0111016;%%
% ``Two-loop superstrings. IV: The cosmological constant and modular forms,''
arXiv:hep-th/0111040.
%%CITATION = HEP-TH 0111040;%%
}

\lref\VV{E.~Verlinde and H.~Verlinde,
``Multiloop Calculations In Covariant Superstring Theory,''
Phys.\ Lett.\ B {\bf 192}, 95 (1987).
%%CITATION = PHLTA,B192,95;%%}
% ``Chiral Bosonization, Determinants And The String Partition Function,''
% Nucl.\ Phys.\ B {\bf 288}, 357 (1987).
%%CITATION = NUPHA,B288,357;%%
}

\lref\FAY{J.D. Fay, ``Theta Functions on Riemann Surfaces'', 
Lecture Notes in Mathematics  Nr. 352, Springer, Heidelberg 1973.}

\lref\BI{M.~Bonini and R.~Jengo,
``Correlation Functions And Zero Modes On Higher Genus Riemann Surfaces,''
Int.\ J.\ Mod.\ Phys.\ A {\bf 3}, 841 (1988).}
%%CITATION = IMPAE,A3,841;%%

\lref\AMV{L.~Alvarez-Gaum\'e, G.W.~Moore and C.~Vafa,
``Theta Functions, Modular Invariance, And Strings,''
Commun.\ Math.\ Phys.\  {\bf 106}, 1 (1986).}
%%CITATION = CMPHA,106,1;%%

\lref\iengoRfour{R.~Iengo,
``Computing the $R^4$ term at two super-string loops,''
JHEP {\bf 0202}, 035 (2002)
[arXiv:hep-th/0202058].
%%CITATION = HEP-TH 0202058;%%
}

\lref\paris{C. Bachas, C. Fabre, E.~Kiritsis, N.A. Obers and P. Vanhove,
``Heterotic/type-I duality and D-brane instantons,''
Nucl.\ Phys.\ B {\bf 509}, 33 (1998)
[arXiv:hep-th/9707126];\br
%%CITATION = HEP-TH 9707126;%%
E. Kiritsis and N.A. Obers,
``Heterotic/type-I duality in $D<10$ dimensions, threshold corrections  and 
D-instantons,''
JHEP {\bf 9710}, 004 (1997)
[arXiv:hep-th/9709058].
%%CITATION = HEP-TH 9709058;%%
}

\lref\LSWF{W. Lerche and S. Stieberger,
``Prepotential, mirror map and F-theory on K3,''
Adv.\ Theor.\ Math.\ Phys.\  {\bf 2}, 1105 (1998)
[Erratum-ibid.\  {\bf 3}, 1199 (1999)]
[arXiv:hep-th/9804176];\br
%%CITATION = HEP-TH 9804176;%%
W. Lerche, S. Stieberger and N.P. Warner,
``Quartic gauge couplings from K3 geometry,''
Adv.\ Theor.\ Math.\ Phys.\  {\bf 3}, 1575 (1999)
[arXiv:hep-th/9811228];
%%CITATION = HEP-TH 9811228;%%
``Prepotentials from symmetric products,''
Adv.\ Theor.\ Math.\ Phys.\  {\bf 3}, 1613 (1999)
[arXiv:hep-th/9901162];\br
%%CITATION = HEP-TH 9901162;%%
K. Foerger and S. Stieberger,
``Higher derivative couplings and heterotic-type I duality in eight  
dimensions,''
Nucl.\ Phys.\ B {\bf 559}, 277 (1999)
[arXiv:hep-th/9901020].
%%CITATION = HEP-TH 9901020;%%
}

\lref\LS{W.~Lerche and S.~Stieberger,
``1/4 BPS states and non-perturbative couplings in N = 4 string theories,''
Adv.\ Theor.\ Math.\ Phys.\  {\bf 3}, 1539 (1999)
[arXiv:hep-th/9907133].
%%CITATION = HEP-TH 9907133;%%
}

\lref\seiberg{M.~Dine and N.~Seiberg,
``Comments on higher derivative operators in some SUSY field theories,''
Phys.\ Lett.\ B {\bf 409}, 239 (1997)
[arXiv:hep-th/9705057];\br
%%CITATION = HEP-TH 9705057;%%
M.~Dine and J.~Gray,
``Non-renormalization theorems for operators with arbitrary numbers of  
derivatives in N = 4 
Yang-Mills theory,''
Phys.\ Lett.\ B {\bf 481}, 427 (2000)
[arXiv:hep-th/9909020].
%%CITATION = HEP-TH 9909020;%%
}

\lref\ssi{S.~Paban, S.~Sethi and M.~Stern,
``Summing up instantons in three-dimensional Yang-Mills theories,''
Adv.\ Theor.\ Math.\ Phys.\  {\bf 3}, 343 (1999)
[arXiv:hep-th/9808119].
%%CITATION = HEP-TH 9808119;%%
}

\lref\SS{S.~Sethi and M.~Stern,
``Supersymmetry and the Yang-Mills effective action at finite N,''
JHEP {\bf 9906}, 004 (1999)
[arXiv:hep-th/9903049].
%%CITATION = HEP-TH 9903049;%%
}

\lref\BPT{I.L. Buchbinder, A.Y. Petrov and A.A.~Tseytlin,
``Two-loop N = 4 super Yang Mills effective action and interaction  between 
D3-branes,''
Nucl.\ Phys.\ B {\bf 621}, 179 (2002)
[arXiv:hep-th/0110173].
%%CITATION = HEP-TH 0110173;%%
}

\lref\MT{R.R.~Metsaev and A.A.~Tseytlin,
``On Loop Corrections To String Theory Effective Actions,''
Nucl.\ Phys.\ B {\bf 298}, 109 (1988).
%%CITATION = NUPHA,B298,109;%%
}

\lref\stieberg{S.~Stieberger,
``(0,2) heterotic gauge couplings and their M-theory origin,''
Nucl.\ Phys.\ B {\bf 541}, 109 (1999)
[arXiv:hep-th/9807124].
%%CITATION = HEP-TH 9807124;%%
}

\lref\brazil{R.~Medina, F.T.~Brandt and F.R.~Machado,
``The open superstring 5-point amplitude revisited,''
JHEP {\bf 0207}, 071 (2002)
[arXiv:hep-th/0208121].
%%CITATION = HEP-TH 0208121;%%
}

\lref\BC{N.~Berkovits and O.~Chandia,
``Massive superstring vertex operator in D = 10 superspace,''
JHEP {\bf 0208}, 040 (2002)
[arXiv:hep-th/0204121].
%%CITATION = HEP-TH 0204121;%%
}

\lref\PSS{S.~Paban, S.~Sethi and M.~Stern,
``Supersymmetry and higher derivative terms in the effective action of 
Yang-Mills theories,''
JHEP {\bf 9806}, 012 (1998)
[arXiv:hep-th/9806028].
%%CITATION = HEP-TH 9806028;%%
}

\lref\belgium{P.~Koerber and A.~Sevrin,
``The non-Abelian Born-Infeld action through order $\alpha'^3$,''
JHEP {\bf 0110}, 003 (2001)
[arXiv:hep-th/0108169].
%%CITATION = HEP-TH 0108169;%%
}

\Title{\vbox{\rightline{HU--EP--02/29} \rightline{NUB--3231}
\rightline{\tt hep-th/0209064}}} {\vbox{\centerline{Non-Abelian
Born-Infeld Action and}
\bigskip\centerline{Type I -- Heterotic Duality (II):}
\smallskip\centerline{Nonrenormalization Theorems}}}
\smallskip
\centerline{S. Stieberger\ and\ T.R. Taylor}
\bigskip
\centerline{ \it Institut f\"ur Physik, Humboldt Universit\"at zu
Berlin,} \centerline{\it Invalidenstra\ss e 110, 10115 Berlin,
FRG} \centerline{\it and} \centerline{\it Department of Physics,
Northeastern University, Boston, MA 02115, USA}

% \medskip
\bigskip\bigskip
\centerline{\bf Abstract} \vskip .2in \noindent Type I --
heterotic duality in $D{=}10$  predicts various relations and
constraints on higher order $F^n$ couplings at different string
loop levels on both sides. We prove the vanishing of two-loop
corrections to the heterotic $F^4$ terms, which is one of the
basic predictions from this duality. Furthermore, we show that the
heterotic $F^5$ and (CP even) $F^6$ couplings are not
renormalized at one loop.
% which imply the absence of those couplings on type I side at
% one-- and two--loop, respectively.
These results strengthen the conjecture that in $D{=}10$ any $\Tr F^{2n}$
coupling appears only at the disk tree-level on type I side and
at $(n{-}1)$-loop level on the  heterotic side. Our
non-renormalization theorems are valid in any heterotic string
vacuum with sixteen supercharges.

\Date{}
\noindent

\newsec{Introduction}

In the recent paper {\STi}, we discussed a class of
higher-derivative $SO(32)$ gauge boson interactions in the
framework of $D{=}10$ type I - heterotic duality. On type I
side, these interactions are related to the non-Abelian
completion of the Born-Infeld action (NBI) which, in more general
context, describes systems of D-branes and orientifold planes. We
considered the NBI action as a power expansion in the gauge field
strength $F$ and examined the $F^6$ interactions in a way
suggested by superstring duality which, for this class of terms,
relates type I at the classical level to the heterotic theory at
two loops. We performed explicit two-loop computations of the
scattering amplitudes and derived the corresponding constraints
on the heterotic $F^6$ interactions. The constraints originate
from Riemann identities reflecting supersymmetry of the
underlying theory, and lead to an unexpected conclusion that the
heterotic $F^6$ terms are not related in any simple way to
Born-Infeld theory. Namely when the gauge bosons are restricted
to the $SO(32)$ Cartan subalgebra generators, the result is
different from the expression obtained by expanding the Abelian
Born-Infeld action. This creates an interesting problem how to
reconcile such a discrepancy with superstring duality.

Type I - heterotic correspondence is a strong-weak coupling
duality so it is guaranteed to work in a straightforward manner
only for quantities that are subject to non-renormalization
theorems. Hence it is very important to investigate possible
corrections to $F^n$ interactions coming from both higher and
lower number of loops. In this paper, we discuss the cases of
$n{=}4,5,6$ in heterotic superstring theory. We derive several
non-renormalization theorems that are directly related to our
discussion of the NBI action. These theorems are also interesting
{\it per se\/} as they apply to the heterotic perturbation theory.

We begin by recalling the relation between the $F^n$ couplings of
type I and heterotic theories. Duality is manifest in the
Einstein frame where the perturbative expansions of the
respective actions can be written as \eqn\einstein{\eqalign{
{\cal S}_{H,\ Einstein}&=\int d^{10}x \sqrt{-g^E_H}\
\lf[R-\fc{1}{4} e^{-\Phi_H/4}\ F^2 -\sum_{n>2}\sum_m\
b_{mn}e^{-m\Phi_H/4}\ F^{n}+\ldots\ri]\cr {\cal S}_{I,\
Einstein}&=\int d^{10}x \sqrt{-g^E_I}\ \lf[R-\fc{1}{4}\
e^{\Phi_I/4}\ F^2 -\sum_{n>2}\sum_m\ b_{mn}e^{m\Phi_I/4}\
F^{n}+\ldots\ri]\ .\cr}} These equations are symbolic in the
sense that $F^{n}$ denotes collectively any Lorentz contraction
and any group theoretical contraction of $n$ gauge field
strengths. The integer $m$ governs the dilaton dependence of each
$F^{n}$ coupling while the corresponding coefficients $b_{mn}$
are constant numbers.\foot{Here, the string tension is normalized
by $2\pi\alpha'=1$ and $\Phi=2\phi$, where $\phi$ is  the standard
dilaton.} In this basis, type I - heterotic duality is manifest
\witten: \eqn\duality{ \Phi_H=-\Phi_I\ \ ,\ \ g_H^E=g_I^E\ \ ,\ \
A_\mu^{a,H}=A_\mu^{a,I}\ .} Transforming \einstein\ into the
string basis with $g_{\mu\nu}^E=e^{-\Phi/4} g_{\mu\nu}^S$ gives
\eqn\strin{\eqalign{ {\cal S}_{H,\ string}&=\int d^{10}x\
\sqrt{-g^S_H}\ \lf[e^{-\Phi_H}R-\fc{1}{4} e^{-\Phi_H}\
F^2-\sum_{n>2}\sum_m\  b_{mn} e^{\Phi_H(n-m-5)/4}\
F^{n}+\ldots\ri]\cr {\cal S}_{I,\ string}&=\int d^{10}x
\sqrt{-g^S_I}\ \lf[e^{-\Phi_I}R-\fc{1}{4} e^{-\Phi_I/2}\
F^2-\sum_{n>2}\sum_m\ b_{mn} e^{\Phi_I(n+m-5)/4}\
F^{n}+\ldots\ri]}} The string loop counting is determined by the
dilaton factor $e^{-\chi\Phi/2}$, with the Euler number
$\chi=2-2g$ for closed strings and $\chi=2-2g-b-c$ for open
strings, where $g$ is the genus of Riemann surface and $b,c$ are
the numbers of holes and crosscaps, respectively. Depending on
the value of $m$, we obtain different chains of $F^{n}$ couplings
at heterotic $(n-m-1)/4$ loops related to the corresponding type I
couplings at order $e^{\Phi_I(n+m-5)/4}$. For instance, the chain
associated to $m=3{-}n$ relates the heterotic $F^{n}$ couplings
at ${n\over 2}{-}1$ loops to type I couplings at the tree level
(disk with $\chi=1$). Changing the value of $m$ by two units
gives other chains. We obtain the following dictionary:

\vskip0.5cm
{\vbox{\ninepoint{
\def\ss#1{{\scriptstyle{#1}}}
$$
\vbox{\offinterlineskip\tabskip=0pt
\halign{\strut\vrule#
%%%%%%%%%%%%%%%%%%
&~$#$~\hfil
% &~$#$~\hfil
&\vrule#&\vrule#
&~$#$~\hfil
&~$#$~\hfil
&~$#$~\hfil
&~$#$~\hfil
&~$#$~\hfil
&~$#$~\hfil
&~$#$~\hfil
&~$#$~\hfil
&~$#$~\hfil
&~$#$~\hfil
&\vrule#&\vrule#
&~$#$~\hfil
&~$#$~\hfil
&\vrule#
\cr
%%%%%%%%%%%%%%%%%%
\noalign{\hrule}
&
{\rm Type\ I}
&&&
\
&&
\
&&
{\rm Heterotic}
&&
\
&&
\
&&&&
\
&&
\cr
%%%%%%%%%%%%%%%%%%
\noalign{\hrule}
&
% {\rm Type\ I}
&&&
{\rm tree~level}
&&
{\rm one~loop}
&&
{\rm two~loops}
&&
{\rm three~loops}
&&
{\rm four~loops}
&&&&
\ \ \ m
&&
\cr
%%%%%%%%%%%%%%%%%%
\noalign{\hrule}
&
\ e^{-\Phi_I/2}
&&&
\ \ \ \ {\bf F^2}
&&
\ \ \ \ {\bf F^4}
&&
\ \ \ \ F^6
&&
\ \ \ \ F^8
&&
\ \ \ \ F^{10}
&&&&
\ 3-n
&&
\cr
%%%%%%%%%%%%%%%%%%
%%%%%%%%%%%%%%%%%%
\noalign{\hrule}
&
\ \ 1
&&&
\ \ \ \ {\bf F^3}
&&
\ \ \ \ F^5
&&
\ \ \ \ F^7
&&
\ \ \ \ F^{9}
&&
\ \ \ \ F^{11}
&&&&
\ 5-n
&&
\cr
%%%%%%%%%%%%%%%%%%
%%%%%%%%%%%%%%%%%%
\noalign{\hrule}
&
\ e^{\Phi_I/2}
&&&
\ \ \ \ {\bf F^4}
&&
\ \ \ \ F^6
&&
\ \ \ \ F^{8}
&&
\ \ \ \ F^{10}
&&
\ \ \ \ F^{12}
&&&&
\ 7-n
&&
\cr
%%%%%%%%%%%%%%%%%%
%%%%%%%%%%%%%%%%%%
%%%%%%%%%%%%%%%%%%
\noalign{\hrule}
&
\ e^{\Phi_I}
&&&
\ \ \ \ F^5
&&
\ \ \ \ F^7
&&
\ \ \ \ F^9
&&
\ \ \ \ F^{11}
&&
\ \ \ \ F^{13}
&&&&
\ 9-n
&&
\cr
%%%%%%%%%%%%%%%%%%
%%%%%%%%%%%%%%%%%%
\noalign{\hrule}
&
\ e^{3/2\Phi_I}
&&&
\ \ \ \ F^6
&&
\ \ \ \ F^8
&&
\ \ \ \ F^{10}
&&
\ \ \ \ F^{12}
&&
\ \ \ \ F^{14}
&&&&
11-n
&&
\cr\noalign{\hrule}}
\hrule}$$
\vskip-10pt
\centerline{\noindent{\bf Table:}
{\sl Possible $F^{n}$ couplings of type I and heterotic string theory in 
$D=10$.
% $e^{\Phi_I(m-5+n)/2}$
}}
% \centerline{\sl following from their duality.}
\vskip10pt}}}

Since the tree-level type I action that originates from the disk
diagram has a single group (Chan-Paton) trace from its boundary,
here we will be mostly interested in this type of group
contractions. These gauge group traces will be denoted by $\Tr$
while the Lorentz group traces by $\tr$. In this case, the couplings
written in the Table in bold face are already known to be
consistent with duality. In particular, $\Tr F^3$ couplings
vanish as a consequence of supersymmetry \GSi. The tree-level
heterotic $\Tr F^4$ couplings are also zero, as shown in \gross.
The heterotic one-loop $\Tr F^4$ has been calculated in
\refs{\LSW,\ellisone} and agrees with type I
\refs{\GSi,\allAbelian} at the tree level. Finally, the absence
of one-loop corrections (from the annulus and M\"obius strip) to
$\Tr F^4$ has been demonstrated in  type I theory in
\refs{\GSii,\BK}.

The first chain, $m=3{-}n$, is particularly interesting. The
$g$-loop heterotic $\Tr F^{2g+2}$ couplings are related by
duality to classical type I theory, hence to the NBI action \AA.
This observation offers a tool for computing the NBI action by
using heterotic perturbation theory. It is quite remarkable that
the tree-level, open string amplitudes are encoded in the
heterotic theory at higher genus. In fact, this motivated us to
perform the two-loop computations in \STi.

The dual actions \strin\ can be compared order by order in
perturbation theory only for quantities that receive contributions
from a limited number of loops. Then some couplings at a given
loop level on one side are simply forbidden because they would
imply a ``negative loop order'' on the dual side. For instance,
the heterotic $\Tr F^4$ couplings must vanish at two loops. In
general, by inspecting the Table, we see that a $\Tr F^{2g+2}$
coupling can appear on the heterotic side only up to the $g$-loop
order, but not beyond. Furthermore, if the NBI action does not
receive corrections beyond the disk level, one would expect that
a $\Tr F^{2g+2}$ term appears on the heterotic side {\it only} at
a $g$-loop order. This would indicate a topological nature of
these couplings, similar to the amplitudes discussed in \AGNT.
However, in view of our findings in \STi, this connection may
hold not for all couplings, but only for those which are BPS
saturated and in some way related to a higher loop generalization of
the elliptic genus \refs{\LSWgenus,\LSW}. In this paper, we prove a number of
perturbative non-renormalization theorems for the heterotic
superstring, all of them consistent with the duality conjecture.
In Section 2, we demonstrate the absence of two-loop corrections
to $F^4$.
In Section 3, we show that all one-loop contributions to $F^5$ are zero.
In Section 4, we extend our one-loop analysis to $F^6$ terms,
which is the case most relevant to \STi. Here again,
we show that all one-loop contributions vanish. In Section 5, we
place these non-renormalization theorems
in a broader context of all-order perturbation theory
and explain their implications for the results of \STi. The paper includes two 
appendices.
In Appendix A, we present some useful tools for computing the one-loop 
amplitudes.
Finally, in the self-contained Appendix B, we show the absence of two-loop 
corrections
to $F^4$ in a formulation which is independent on the choice of gauge slice.

\newsec{Vanishing of heterotic two-loop $F^4$}
% \goodbreak

The heterotic $\Tr F^4$ terms have been calculated at one loop in
$D=10$ in \refs{\LSW,\ellisone}. It has been speculated for a
long time that the two-loop correction to $\Tr F^4$ may be absent,
because $t_8 \Tr F^4$ appears in the same superinvariant as the
Green-Schwarz anomaly term $\epsilon_{10}B \Tr F^4$. According to
Adler-Bardeen theorem, anomalies are not renormalized: at two
loops, this has been shown explicitly for Green-Schwarz anomaly
in \yasuda.

In this section we will prove that the two-loop correction to $\Tr
F^4$ does indeed vanish. To that end, we consider the correlator
of four gauge bosons:
 \eqn\start{ \vev{
V_{A^{a_1}_{\mu_1}}(z_1,\ov z_1) V_{A^{a_2}_{\mu_2}}(z_2,\ov z_2)
V_{A^{a_3}_{\mu_3}}(z_3,\ov z_3) V_{A^{a_4}_{\mu_4}}(z_4,\ov z_4)
Y(x_1) Y(x_2) }\ ,}with the corresponding vertex operators taken
in the zero-ghost picture, \eqn\vertex{ V_{A^a_\mu}(z,\ov z)
=\epsilon_\mu\ :J^a(\ov z)\ \lf[\p
X^\mu+i(k_\nu\psi^\nu)\psi^\mu\ri]\ e^{ik_\rho X^\rho(z,\ov z)}:}
and with two necessary picture changing operator (PCO) insertions $Y$ at 
arbitrary points
$x_1$ and $x_2$, see \STi. The gauge currents can be fermionized
as: \eqn\gauge{ J^a(z)=(T^a)_{ij}\ \psi^i(z)\psi^j(z)\ ,} where
$T^a$ are the $SO(32)$ gauge group generators  in the defining
representation. The space-time part of this amplitude has been
already discussed in \refs{\iengo,\LP}. In our proof, we will
combine it with the gauge part.

Essentially, there are two different ways to tackle two-loop
calculations. One way is the so-called $\theta$-function
approach, where the partition function and correlators are
expressed in terms of genus two $\theta$-functions. After choosing the unitary 
gauge,\foot{The unitary gauge is a special choice of the PCO insertion points. 
Different choices are related by total derivatives w.r.t.\ the moduli of
the two-loop Riemann 
surface \VV. These contributions are zero provided these derivative terms 
vanish 
at the boundaries of the moduli space. This has to be verified for each 
amplitude.
A very useful approach to handle these complications has been 
recently elaborated by D'Hoker and Phong \HP. In Appendix \appB\ we will 
address the 
vanishing of two-loop $\Tr F^4$ within that framework.}
this method proves to be very useful {\it e.g}.\ for identifying the 
combinations of
amplitudes that vanish due to Riemann identities, as demonstrated
in \STi. 
The other way, which is more suitable for the present
discussion, uses the hyperelliptic formalism. This approach
allows a more transparent treatment of the ambiguity in choosing
the gauge slices, {\it i.e}.\ fixing the PCO positions.
In the hyperelliptic formalism, the genus two surface
is represented by a two-sheet covering of the complex plane,
\eqn\curve{ y(z)^2=\prod_{i=1}^6 (z-a_i)\ ,} with six branch
(ramification) points $a_i$. The string correlators are functions
on this hyperelliptic surface. We refer the reader to
\refs{\iengo,\zhu} and references therein for more information.

Let us introduce the basic ingredients of hyperelliptic
formalism. The space-time zero mode contribution $\det \im \Om$
is related to the corresponding quantity $T$ on the hyperelliptic
surface:
\eqn\T{ T(a_i,\ov a_i)=\int d^2u
d^2v\ \fc{|u-v|^2}{|y(u)y(v)|^2}=2|\det K|^2 \det\im\Om\ .}
Here, $\Omega$ are the moduli of the genus two Riemann surface, defined
by integrals $\Omega_{ij}=\oint_{b_j} \omega_i$ of the canonical one-forms $\omega_i$
over the $b$-cycles $b_j$. The
determinant factor $|\det K|^2$ arises because we are working with the 
non-canonically
normalized Abelian differentials $\tilde\omega_i=\omega_jK_{ji}$,
with $\tilde \omega_1(z)=\fc{dz}{y(z)}$ and $\tilde
\omega_2(z)=\fc{z\ dz}{y(z)}$, respectively. On the
hyperelliptic Riemann surface, even spin structures $\delta$ are
in one-to-one correspondence with the splittings of six branch
points $a_i$ into two non-intersecting sets $\{A_1,A_2,A_3\}$ and
$\{B_1,B_2,B_3\}$. The chiral fermion determinant is given by
\eqn\chiraldet{ \det_\delta\p_{1/2}= \alpha^{-1}\prod_{i<j}
\lf(A_i-A_j\ri)^{1/4}\lf(B_i-B_j\ri)^{1/4}\equiv\fc{Q_\delta^{1/4}}{\alpha}\
,} where $\alpha$ represents the oscillator contributions:
\eqn\osci{ \alpha=\prod_{i<j}(a_i-a_j)^{1/8}\ .}
The quantity $Q_\delta$ is related to the familiar
genus two $\theta$-functions $\th_\delta(0,\Om)$ through the
Thomae formula \mumford: $\th_\delta(0,\Om)=(\det
K)^{1/2}Q_\delta^{1/4}$. For even spin structure $\delta$, the
Szeg\"o kernel takes the form \morozov: \eqn\szego{
\vev{\psi(z_1)\psi(z_2)}_\delta=\h\fc{dz_1^{1/2}dz_2^{1/2}}{z_1-z_2}
\fc{u_\delta(z_1)+u_\delta(z_2)}{\sqrt{u_\delta(z_1)u_\delta(z_2)}}\
.} Here, the functions $u_{\delta}(z)$ are introduced as
\eqn\uz{
u_{\delta}(z)=\fc{(z-A_1)(z-A_2)(z-A_3)}{y(z)}\ ,} where $A_i$ are the
three branch points $a_i$ associated to a given spin structure
$\delta$. Finally, the (three-dimensional) genus two measure
$d\mu$ can be expressed in terms of integrals over three branch
points: \eqn\measure{
d\mu=\fc{d^2a_1d^2a_2d^2a_3|a_{45}a_{56}a_{46}|^2}{
\prod\limits_{i<j}^6|a_{ij}|^2}\ .}

The space-time part of the two-loop amplitude under consideration
has been previously evaluated within the hyperelliptic formalism
by several authors \iengo. The result is: \eqn\their{
\triangle_{t_8\Tr F^4}^{\rm 2-loop} =\int d\mu\ T^{-5}\
\prod_{l=1}^4\fc{d^2z_l(x-z_l)}{y(z_l)}\ I(x)\ F(\va;\vz)\ .}
Here, $I(x)$ summarizes all contributions from PCOs whose two
positions have been chosen at the same point $x$ on the upper and
lower sheets. The amplitude is independent of this choice, as any
change amounts to a total derivative on the moduli space \VV. The
kinematics of the four gauge boson amplitude corresponds to the
familiar $t_8$-tensor. All eight space-time fermions from
\vertex\ are contracted in \their. This gives already $\Oc(k^4)$
in momenta. Fewer fermion contractions would give a vanishing
result. The expression \their\ assumes zero momenta in the
exponentials of the gauge vertices \vertex\ as they would bring
down more momentum factors. This step needs some care because
there may be potential singularities which would decrease the
power of momentum. We shall give a justification of this step
later. Finally, the function $F(\va;\vz)$ encodes the gauge part
from the right-moving sector. It comprises the $SO(32)$ gauge
partition function (cf. Eqs. \chiraldet\ and \osci)
\eqn\gaugelattice{ Z_{SO(32)}(a)=\alpha^{-16}\sum_{\beta\ {\rm
even}}Q_\beta^4\ ,} supplemented with the four gauge current
correlator
$\vev{J^{a_i}(z_i)J^{a_j}(z_j)J^{a_k}(z_k)J^{a_l}(z_l)}$ that
we shall determine now.

In order to obtain the group theoretical structure
$\Tr(T^{a_i}T^{a_j}T^{a_k}T^{a_l})$, the gauge fermions of \gauge\
have to be contracted in such a way that their four vertex
positions $z_i$ form a closed loop (square): \eqn\Gauge{\eqalign{
B_{\bet}(z_i,z_j,z_k,z_l)&\equiv
\vev{\psi(z_i)\psi(z_j)}_{\bet}\vev{\psi(z_j)\psi(z_k)}_{\bet}
\vev{\psi(z_k)\psi(z_l)}_{\bet}\vev{\psi(z_l)\psi(z_i)}_{\bet}\cr
&=\fc{1}{16}\fc{dz_1dz_2dz_3dz_4}{z_{ij}z_{jk}z_{kl}z_{li}}\
\fc{1}{u_\bet(z_i)u_\bet(z_j)u_\bet(z_k)u_\bet(z_l)}\cr &\times
[u_\bet(z_i)+u_\bet(z_j)][u_\bet(z_j)+u_\bet(z_k)][u_\bet(z_k)+u_\bet(z_l)]
[u_\bet(z_l)+u_\bet(z_i)]\ .}} Due  to the symmetry of the
space-time part \their, it is sufficient to focus on one specific
contraction, say $B(z_1,z_2,z_3,z_4)$. However, it is more
convenient to take the combination ${1\over 3}[
B(z_1,z_2,z_3,z_4)+B(z_1,z_2,z_4,z_3)+B(z_1,z_3,z_4,z_2)]$ which
shows a particularly simple dependence on the vertex positions
$z_i$: \eqn\combination{\eqalign{ &{1\over 3}[
B_{\bet}(z_1,z_2,z_3,z_4)+B_{\bet}(z_1,z_2,z_4,z_3)+B_{\bet}(z_1,z_3,z_4,z_2)]\cr
&=\fc{1}{48} \fc{dz_1dz_2dz_3dz_4}{y(z_1)y(z_2)y(z_3)y(z_4)}\
\lf[c_0(a_i)+c_1(a_i)\sum_i z_i+ c_2(a_i)\sum_{i<j} z_iz_j\ri.\cr
&\hskip3cm\lf.+c_3(a_i)\sum_{i<j<k} z_iz_jz_k+c_4(a_i)
z_1z_2z_3z_4\ri]\ ,}} with the spin structure dependent
coefficients $c_j(a_i)$ being polynomials in $a_i$. Their explicit form
is not important for our arguments.
{}Furthermore, for later use, we note that the bracket of
\combination\ scales with the weight $\lambda^8$ under the
simultaneous rescalings $a_i\ra\lambda a_i$ and $z_i\ra\lambda
z_i$. The r.h.s.\ of \combination\  involves linear combination of
two Abelian differentials $\fc{dz}{y(z)}$ and $\fc{z\ dz}{y(z)}$,
introduced earlier. The above equation represents one of
identities that can be found in \FAY, rewritten in the
hyperelliptic formalism. The total gauge part $F(\va;\vz)$ from
\their\ is obtained by combining \combination\ with the $SO(32)$
lattice sum \gaugelattice: \eqn\totalgauge{\eqalign{
F(\vaw;\vzw)&={1\o
48}\fc{\alpha^{-16}}{y(z_1)y(z_2)y(z_3)y(z_4)}\times\sum_\beta\
Q_\beta^4\ [c_0(a_i)+c_1(a_i)\sum_i z_i+\cr
&\hiii+c_2(a_i)\sum_{i<j} z_iz_j+c_3(a_i)\sum_{i<j<k} z_iz_jz_k+
c_4(a_i) z_1z_2z_3z_4]\ .}} This expression is particular useful,
since in this form all possible singularities at $z_i\ra z_j$
are eliminated. This justifies setting the momenta of the
exponentials \vertex\ to zero at the beginning.

By following the same steps as in \JZ, using simultaneous
$SL(2,\IC)$ transformations on $a_i,z_i$ and $x$, the expression
\their\ can be rewritten\foot{We call the $SL(2,\IC)$ transformed
values again $a_i,z_i$ and $x$.} in an explicitly modular
invariant form: \eqn\hyper{ \triangle_{t_8\Tr F^4}^{\rm
2-loop}=\int_\IC d\tilde\mu\ \prod_{l=1}^4\fc{d^2
z_l(x-z_l)}{|y(z_l)|^2}\ \prod_{i=1}^3\delta^2(z_i-z_i^0)\
\prod_{j<k}^3|z_j^0-z_k^0|\ I_M(x)\ \tilde F(\va;\vz)\ ,} where
\eqn\ftilde{\tilde
F(\vaw;\vzw)=y(z_1)y(z_2)y(z_3)y(z_4)\ F(\vaw;\vzw)\ ,} and the
measure is \eqn\measure{
d\tilde\mu=\fc{\prod\limits_{i=1}^6d^2a_i}{T^5\prod\limits_{i<j}^6
|a_i-a_j|^2}\ .} Modular invariance manifests in \hyper\ as an
invariance under the permutations of $a_i$. Now, three vertex
positions $z_i$ are fixed to $z_i^0$ ($i=1,2,3$) at the cost of
allowing for integration over all six branch points $a_i$ rather
than three. The function $I_M(x)$ denotes the PCO contributions
\JZ: \eqn\IM{\eqalign{
I_M(x)&=\fc{1}{4}\sum_{i=1}^6\fc{1}{(x-a_i)^2}-
\fc{1}{4}\sum_{i<j}^6\fc{1}{x-a_i}\fc{1}{x-a_j}-\fc{1}{8}\sum_{i=1}^6\fc{1}{x-a_i}
\sum_{l=1}^4\fc{1}{x-z_l}\cr &+{1\o
4}\sum_{k<j=1}^4\fc{1}{x-z_k}\fc{1}{x-z_j}-\fc{5}{4} \sum_{i=1}^6
\fc{1}{x-a_i} \fc{\p}{\p a_i}\ln T\ .}} The amplitude
$\triangle_{t_8\Tr F^4}^{\rm 2-loop}$ is independent of the point
$x$, which is the insertion point of the two PCOs. Any
change in $x$ amounts to a total derivative in moduli space \JZ.

A great simplification occurs when we choose $x$ equal to the
vertex position $z_1$ ($x\ra z_1^0$): \eqn\hyperi{
\triangle_{t_8\Tr F^4}^{\rm 2-loop}=V\int_\IC d\tilde\mu\int_\IC
d^2z\ \fc{z^0_1-z}{|y(z_1^0)y(z^0_2)y(z^0_3)y(z)|^2}\ I_\infty\
\tilde F(\va;\{\ov z_n^0,\ov z\})\ ,} with \eqn\VI{\eqalign{
I_\infty&=-\fc{1}{8}\sum_{i=1}^6\fc{1}{z_1^0-a_i}+\fc{1}{4}
\fc{1}{z_1^0-z_2^0}+\fc{1}{4}\fc{1}{z_1^0-z_3^0}+\fc{1}{4}\fc{1}{z_1^0-z}\ ,\cr
V&=(z^0_1-z^0_2)^2 (z^0_1-z^0_3)^2 (\ov z^0_1-\ov
z^0_2)|z^0_2-z^0_3|^2 (\ov z^0_1-\ov z^0_3)\ .}} In the limit
$z\ra z_1^0$ the integrand in \hyperi\ remains finite. However,
it is not clear what happens when $a_i\ra z_1^0$. This limit can
be analyzed\foot{The gauge function $\tilde F(\va,\vz)$  becomes
at most zero for some or all $a_i$ approaching $z_1^0$.}  by inspecting a
potentially more singular expression, \eqn\inspection{ \int_\IC
d^2\tilde\mu \int_{\IC} \fc{d^2z}{|y(z)|^2}\
\fc{1}{|y(z_1^0)y(z_2^0)y(z_3^0)|^2}\ .} This is the same
expression that appears in the context type IIA two-loop $R^4$
terms, for which the finiteness of \inspection\ has been verified
in the limit $a_i\ra z_j^0$ \JZ.

The integrand of \hyperi\ is invariant under the modular
transformation \eqn\modular{ z_i^0\ra\fc{az_i^0+b}{cz_i^0+d}\ \
,\ \ ad-bc=1\ .} This allows to fix three positions. A convenient
choice is: \eqn\fix{ z_1^0=0\ \ ,\ \ z_2^0=\infty\ \ ,\ \
z_3^0=x\neq 0,\ \infty\ .} With this choice, Eq.\hyperi\ becomes:
\eqn\hyperfix{ \triangle_{t_8\Tr F^4}^{\rm 2-loop}=-{1\o
4}x|x|^2\int_\IC d\tilde\mu \int_\IC d^2z\fc{z}{|y(z)y(0)y(x)|^2}\
\lf(\h\sum_{i=1}^6\fc{1}{a_i}-\fc{1}{x}-\fc{1}{z} \ri)\
F_\infty(\va;\ov x,\ov z)\ ,} where \eqn\introf{
F_\infty(\vaw;x,z)\equiv\lf.\fc{\p}{\p z_2} \tilde
F(\vaw;\vzw)\ri|_{{z_1^0=0,\ z_3^0=x,\atop z_4=z}}\ .} Observing
$\p_{a_i} y(0)^{-1}=-\h a_i^{-1}y(0)^{-1}\ , \sum_i\p_{a_i}
y(x)^{-1}=\sum_i\h (x-a_i)^{-1}y(x)^{-1}=-\p_x y(x)^{-1}$, allows
us to rewrite $\triangle_{t_8\Tr F^4}^{\rm 2-loop}$ in a more
convenient form after partial integrations: \eqn\conv{
\triangle_{t_8\Tr F^4}^{\rm 2-loop}={1\o 4}|x|^2\lf(x\fc{\p}{\p
x}+1\ri) \int_\IC d\tilde\mu\int_\IC
d^2z\fc{z}{|y(z)y(0)y(x)|^2}\ F_\infty(\va;\ov x,\ov z)\ .}
The whole expression \conv\ (or \hyperfix) is  independent of $x$.
This can be verified\foot{The following steps are similar to those of Ref.
\iengoRfour\ demonstrating the vanishing of two-loop corrections to the 
$t_8t_8 \Tr R^4$
term in type IIA/B theories.}  by allowing the changes $x\ra\lambda x,\
z\ra \lambda z$ and $a_i\ra \lambda a_i$ under which the
transformation rules $T\ra \lambda^{-3}\ov\lambda^{-3}T$,
$d\tilde\mu\ra \lambda^6\ov\lambda^6d\tilde\mu$, and
$F_\infty(\{\lambda a_m \};\lambda x,\lambda z)\ra \lambda
F_\infty(\vaw; x, z)$ follow. Therefore in the integrand of
\conv\ we may choose $x=1$:
\eqn\convnew{ \triangle_{t_8\Tr F^4}^{\rm
2-loop}={1\o 4}|x|^2\lf(x\fc{\p}{\p x}+1\ri) |x|^{-2}\int_\IC
d\tilde\mu\int_\IC d^2z\fc{z}{|y(z)y(0)y(1)|^2}\ F_\infty(\va;1,\ov z)\ .}
However, due to $(x\fc{\p}{\p x}+1)x^{-1}=0$, we
conclude: \eqn\finalfour{ \triangle_{t_8\Tr F^4}^{\rm 2-loop}=0\ .}

For the group contraction $(\Tr F^2)^2$, instead of \totalgauge,
the relevant gauge part is:
\eqn\relevgauge{ F(\vaw,\vzw)={1\o
16}\fc{\alpha^{-16}}{(z_1-z_2)^2(z_3-z_4)^2}\sum_\bet Q_\bet^4\
\fc{[u_\bet(z_1)+u_\bet(z_2)][u_\bet(z_3)+u_\bet(z_4)]}{u_\bet(z_1)u_\bet(z_2)
u_\bet(z_3)u_\bet(z_4)}\ .}
Again, $\tilde F(\vaw,\vzw)$ (defined through \ftilde),
which enters\foot{The poles appearing in the integrand \hyper\ in the limit
for $z_i\ra z_j$ can be analytically continued to a finite value.}
\hyper, shows the previously encountered behaviour.
Namely its $z_2$-independent part (defined by \introf)
transforms as $F_\infty(\{\lambda a_m\};\lambda x,\lambda z)\ra \lambda
F_\infty(\vaw; x, z)$ under the rescalings $x\ra\lambda x,\
z\ra \lambda z$ and $a_i\ra \lambda a_i$.
Aside from these properties of the gauge
part, the essential steps to prove \finalfour\ affected only the
space-time part. Therefore, we derive also for the group
contraction $(\Tr F^2)^2$: \eqn\finaltwoloop{ \triangle_{t_8(\Tr
F^2)^2}^{\rm 2-loop}=0\ .} Thus we have established the two-loop
non-renormalization theorems \finalfour\ and \finaltwoloop\ 
in $D{=}10$  heterotic string theory.

% For toroidal compactifications
% these calculation essentially do not change and our
% renormalization theorem, proven in ten dimensions, hold in any
% dimension with sixteen supercharges.
\newsec{Vanishing of heterotic one-loop $F^5$}
% \goodbreak

In the past, only 1/2 BPS-saturated one-loop amplitudes have been
discussed\foot{With the exception of Ref.\LS, where also 1/4 BPS saturated
amplitudes have been considered.} in heterotic string vacua with
sixteen supercharges. They describe couplings which are related
by supersymmetry to eight-fermion terms. Their characteristic
feature is that they depend on the ground state only of the
right-moving sector. The latter then contributes just as a
constant, and one is left with world-sheet torus integrals over
anti-holomorphic functions representing the contributions of the
left-moving sector. This will no longer be the case once we
consider non 1/2 BPS-saturated amplitudes involving more than
eight fermions, which generically receive also non-constant
contributions from the right-moving sector.

In this section, we prove that the one-loop corrections to the two
possible $F^5$ space-time contractions, $\tr F^5$ (denoted by
$P$) and $\tr F^3 \tr F^2$ (denoted by $S$), vanish exactly for
any gauge contraction. To that end, we consider the five-point
gauge boson amplitude, \eqn\SMfive{
\vev{V_{A^{a_1}_{\mu_1}}(z_1,\ov z_1)\ldots
V_{A^{a_5}_{\mu_5}}(z_5,\ov z_5)}_{\rm even}} and extract the
relevant kinematic pieces. The gauge boson vertex operator (in
the zero-ghost picture) is given in \vertex\ and the gauge
currents are fermionized according to \gauge. The one-loop
fermion propagator for even spin structure
$\vec\alpha=(\al_1,\al_2)$ is \eqn\green{ G_{\vec\al}^F(z_{12})\
\delta^{\mu\nu}= \vev{\psi^\mu(z_1)\psi^\nu(z_2)}_{\vec\al}= \fc{
\th_{\vec\al}(z_{12},\tau)\ \th'_1(0,\tau)}{\th_1(z_{12},\tau)\
\th_{\vec\al}(0,\tau)}\ \delta^{\mu\nu}\ .} In the following, we
shall first discuss the gauge pentagon, $\Tr F^5$ case. In order
to yield the corresponding group theoretical factor
$\Tr(T^{a_i}T^{a_j}T^{a_k}T^{a_l} T^{a_m})$, the world-sheet
gauge fermions \gauge\ must be contracted in such a way that the
vertex positions $z_i$ form a pentagon. Hence the gauge part for
the spin structure $\vec\be=(\be_1,\be_2)$ becomes \eqn\pentagon{
f_{\vec\be}(\ov q,\vz)= G_{\vec\be}^F(\ov z_{ij})\
G_{\vec\be}^F(\ov z_{jk})\ G_{\vec\be}^F(\ov z_{kl})\
G_{\vec\be}^F(\ov z_{lm})\ G_{\vec\be}^F(\ov z_{mi})\ .} Due to
the periodicity properties of the fermion propagator,
$G_{\vec\bet}^F(z+1,q)=-e^{2\pi i\bet_1}G_{\vec\bet}^F(z,q)$ and
$G_{\vec\bet}^F(z+\tau,q)=-e^{-2\pi i\bet_2}G_{\vec\bet}^F(z,q)$,
we conclude that the expression \pentagon\ is periodic under the
transformations $z_i\ra z_i+1$ and $z_i\ra z_i+\tau$.

Space-time supersymmetry requires that at least eight world-sheet
fermions from the five vertex operators \vertex\ are taken into
account. Each vertex operator supplies a pair of fermions at the
same position $z_i$. Therefore we have to contract all ten
fermions. This gives already the required $\Oc(k^5)$ order in
momentum and we may set the momenta of the exponentials in
\gauge\ to zero. We shall comment on this step at the end of this
section. Two space-time kinematics\foot{Due to kinematical
reasons, there are no CP odd $F^5$ couplings in $D=10$.} $\Kc_P$
and $\Kc_S$ are possible, depending on how the ten fermions are
contracted:\foot{Diagramatically, the corresponding Lorentz contractions can 
be represented by a pentagon ($P$) and by a triangle plus one line ($S$), 
respectively.} \eqn\spacev{\eqalign{
g_{\vec\alpha}^{\Kc_S}(q,\vzw)&=-G^F_{\vec\alpha}(z_{12})\
G^F(z_{23}) \ G^F_{\vec\alpha}(z_{31}) \
G^F_{\vec\alpha}(z_{45})^2\ ,\cr g_{\vec\alpha}^{\Kc_P}(q,\vzw)&=
-G_{\vec\alpha}^F(z_{12})\ G_{\vec\alpha}^F(z_{23})\
G_{\vec\alpha}^F(z_{34})\ G_{\vec\alpha}^F(z_{45})\
G_{\vec\alpha}^F(z_{51})\ . }} Thus in total, the two different
kinematics $\Kc$ receive the following one-loop corrections:
\eqn\resultv{ \Delta^{\rm 1-loop}_{\Kc\Tr
F^5}=\int\limits_\Fc\fc{d^2\tau}{\tau_2^6} \fc{1}{\eta(\ov
q)^{24}}\fc{1}{\eta(q)^{12}}\ \ \sum_{\vec\alpha,\vec\be}
s_{\vec\al}\ \th_{\vec\alpha}(q)^4 \theta_{\vec\be}(\ov q)^{16}\
\int\limits_{\Ic_\tau}\prod_{i=1}^5\  d^2 z_i \
g_{\vec\alpha}^\Kc(q,\vzw)\ f_{\vec\be}(\ov q,\vz)\ .} Here, the
sum over $\vec\al$ represents the even spin structure sum (with
the phases $s_{\vec\al}=(-1)^{2\al_1+2\al_2}$) and the sum over
even $\vec\bet$ is the $SO(32)$ gauge lattice sum. In general,
the world-sheet torus integrals (with the integration region
$\Ic_\tau=\{z\ |\ -\h\leq\re(z)\leq\h\ ,\
0\leq\im(z)\leq\tau_2\}$) over the five positions imply a
coupling between the left-moving $f_{\vec\be}(\ov q,\vz)$ and
right-moving $g_{\vec\alpha}^{\Kc}(q,\vzw)$ parts, which
complicates the procedure. In fact, so far, only purely
antiholomorphic, $z_i$-independent world-sheet torus integrals
have been discussed in the literature. This is the case when the
amplitude represents a BPS-saturated coupling. Then the right--moving 
sector is in the ground state and contributes a constant
to the full amplitude, without a holomorphic position
$z_i$-dependence.

After recalling the relation of the square of the fermionic
propagator \eqn\identity{ G^F_{\vec\al}(z)^2= \lf(\fc{
\th_{\vec\al}(z,\tau)\ \th'_1(0,\tau)}{\th_1(z,\tau)\
\th_{\vec\al}(0,\tau)}\ri)^2 =\ \fc{\p^2}{\p z^2}  \ln
\th_{\vec\al}(0,\tau)+\fc{\pi}{\tau_2}-\p^2 G_B(z)\ ,} with the
bosonic Greens function $G_B$ \eqn\Bosonic{ \p
G_B(z)=\p\ln\th_1(z,\tau)+\fc{2\pi i}{\tau_2}\im(z)\ ,} we
proceed to the evaluation of the spin structure sum $\vec\al$ in
\resultv. We obtain\foot{The whole right-moving part of the
integrand \resultv\ is put into $\tilde g$.} \eqn\spinv{\eqalign{
\tilde g^{\Kc_S}(q,\vzw)&=-\fc{(-2\pi)^3}{\eta^3}\sum_{\vec\al}
s_{\vec\al} \fc{\p^2}{\p z^2}\th_{\vec\al}(0,\tau)\
\fc{\th_{\vec\al}(z_{12})}{\th_1(z_{12})}\fc{\th_{\vec\al}(z_{23})}{\th_1(z_{23})}
\fc{\th_{\vec\al}(z_{31})}{\th_1(z_{31})}\ ,\cr \tilde
g^{\Kc_P}(q,\vzw)&=-(-2\pi)^5\eta^3\sum_{\vec\al}
s_{\vec\al}\fc{1}{\th_{\vec\al}(0)}\
\fc{\th_{\vec\al}(z_{12})}{\th_1(z_{12})}\fc{\th_{\vec\al}(z_{23})}{\th_1(z_{23})}
\fc{\th_{\vec\al}(z_{34})}{\th_1(z_{34})}\fc{\th_{\vec\al}(z_{45})}{\th_1(z_{45})}
\fc{\th_{\vec\al}(z_{51})}{\th_1(z_{51})}\ ,}} for the kinematics
$\Kc_S$ and $\Kc_P$, respectively. Since the last term $\p^2
G_B(z)$ of \identity\ is a periodic function and the gauge part
$f_{\vec\bet}(\ov q,\vz)$ is also periodic, it will give a
vanishing contribution to \resultv\ after performing the position
integral: \eqn\vanishbos{ \int_{\Ic_\tau}d^2z_jd^2z_k\
\p^2G_B(z_j-z_k)\ f_{\vec\be}(\ov q,\vz)=0\ .} This is why we
dropped that term in $\tilde g^{\Kc_S}(q,\vzw)$. We shall
simplify the sums \spinv\ in the appendix \appAiv\ by a combined
action of Riemann and Fay trisecant identities. These
manipulations result in the following expression for $\Delta^{\rm
1-loop}_{\Kc\Tr F^5}$ \eqn\resultvv{ \Delta^{\rm 1-loop}_{\Kc\Tr
F^5}=\int\limits_\Fc\fc{d^2\tau}{\tau_2^6}
 \sum_{\vec\be}\ \fc{\theta_{\vec\be}(\ov q)^{16}}{\eta(\ov q)^{24}}
\int\limits_{\Ic_\tau}\prod_{i=1}^5\  d^2 z_i \ \tilde
g^\Kc(q,\vzw)\ f_{\vec\be}(\ov q,\vz)\ ,} with:
\eqn\SpinVa{\eqalign{ \tilde g^{\Kc_S}(q,\vzw)&=-(2\pi)^4\ [\p
G_B(z_1-z_2)+\p G_B(z_2-z_3)+\p G_B(z_3-z_1)]\ ,\cr \tilde
g^{\Kc_P}(q,\vzw)&=(2\pi)^4\ [\p G_B(z_1-z_2)+\p G_B(z_2-z_3)+\p
G_B(z_3-z_4)+\p G_B(z_4-z_5)\cr 
&\hiii +\p G_B(z_5-z_1)]\ .}}
This form of the spin structure-dependent piece of \resultv\ is
very convenient for performing the integrations over the
positions $z_i$. Indeed, since the gauge part $f_{\vec\bet}(\ov
q,\vz)$ is periodic at the boundary of $\Ic_\tau$, the integral
\resultvv\ vanishes after partial integration: \eqn\intvan{
\int_{\Ic_\tau}\ d^2z_j\ \p G_B(z_j-z_k)\ f_{\vec\bet}(\ov
q,\vz)=0\ .} We conclude: \eqn\FINALv{ \Delta^{\rm
1-loop}_{\Kc_{S,P}\Tr F^5}=0\ .} Since it is the form of the
space-time part \SpinVa\ which leads to the vanishing of $\Tr F^5$
couplings, we may conclude the same for the other
group-theoretical contraction: $\Delta^{\rm 1-loop}_{\Kc_{S,P}\Tr
F^2\Tr F^3}=0$.

The form of \SpinVa\ is very useful for analyzing eventual
singularities that could appear when the vertex positions
$z_i,z_j$ approach each other. Due to supersymmetry, there are no
singularities from three, four or five points colliding. However,
the so-called pinch effects appear in certain regions of the
integration domain $\Ic_\tau$ in the limit $z_i\ra z_j$. Then the
momenta of the exponentials of \vertex\ cannot be {\it a priori}
neglected. In this limit, the fermionic correlators behave as
$G_{\vec\alpha}(z_{ij})\ra 1/z_{ij}, G_{\vec\bet}(\ov z_{ij})\ra
1/\ov z_{ij}$, while the exponentials $\vev{e^{ik_i X(z_i,\ov
z_i)}e^{ik_j X(z_j,\ov z_j)}}\sim |z_{ij}|^{\alpha'k_1 k_2}$, and
we encounter: \eqn\pinch{ \int_{|z_{ij}|<\epsilon}
\fc{|z_{ij}|^{\al'k_i k_j}}{|z_{ij}|^2}= \fc{2\pi}{\alpha'k_i
k_j}\ .} This signals poles from massless particle exchanges in
one-particle reducible diagrams \PINCH. Setting the momenta of
the exponentials to zero means that we neglect such reducible
contributions, which is the right thing to do when discussing the
effective action.

\newsec{Vanishing of heterotic one-loop $F^6$}

In this section, we prove that the one-loop corrections to the (CP
even) space-time kinematics, $\tr F^6,\ \tr F^4 \tr F^2$, $(\tr
F^2)^3$ and $(\tr F^3)^2$, vanish exactly for any gauge
configuration. These four space-time contractions will be denoted
by $H$, $S$, $L$ and $T$, respectively, referring to their
diagrammatic representation, see \STi. We will consider the
six-point gauge boson amplitude: \eqn\SM{
\vev{V_{A^{a_1}_{\mu_1}}(z_1,\ov z_1)\ldots
V_{A^{a_6}_{\mu_6}}(z_6,\ov z_6)}_{\rm even}} and extract the
relevant kinematic pieces. The gauge boson vertex operator is
given in \vertex\ and the gauge currents are fermionized
according to \gauge. The one-loop fermion propagator is written
for even spin structure in \green. In the following, we shall
first discuss the gauge hexagon $\Tr F^6$ case. In order to yield
the corresponding group-theoretical factor
$\Tr(T^{a_i}T^{a_j}T^{a_k}T^{a_l} T^{a_m}T^{a_n})$, the
world-sheet gauge fermions \gauge\ must be contracted in such a
way that the vertex positions $z_i$ form a hexagon. Thus for the
spin structure $\vec\be=(\be_1,\be_2)$, the gauge part becomes
\eqn\hex{ f_{\vec\be}(\ov q,\vz)= G_{\vec\be}^F(\ov z_{ij})\
G_{\vec\be}^F(\ov z_{jk})\ G_{\vec\be}^F(\ov z_{kl})\
G_{\vec\be}^F(\ov z_{lm})\ G_{\vec\be}^F(\ov z_{mn})\
G_{\vec\be}^F(\ov z_{ni})\ .}

Space-time supersymmetry requires that at least eight fermions
from the six vertex operators \vertex\ are taken into account.
Thus, we have to consider two cases: contracting eight fermions
or all twelve fermions. Let us discuss the latter case first.
Then the respective parts of vertices yield the desired $\Oc(k^6)$
order in momentum. Thus we may neglect\foot{See also the comment
made at the end of the previous section.} the exponentials as
they would increase the power of momentum. Depending on the way
how these twelve fermions are contracted, four different
kinematical configurations $\Kc_L,\Kc_S,\Kc_H$ and $\Kc_T$ arise.
They correspond to the following contractions:
\eqn\space{\eqalign{
g_{\vec\alpha}^{\Kc_L}(q,\vzw)&=G^F_{\vec\alpha}(z_{12})^2\
G^F_{\vec\alpha}(z_{34})^2\ G^F_{\vec\alpha}(z_{56})^2\ ,\cr
g_{\vec\alpha}^{\Kc_S}(q,\vzw)&=-G^F_{\vec\alpha}(z_{12})\
G^F(z_{23}) \ G^F_{\vec\alpha}(z_{34}) \ G^F_{\vec\alpha}(z_{41})\
G^F_{\vec\alpha}(z_{56})^2\ ,\cr g_{\vec\alpha}^{\Kc_H}(q,\vzw)&=
-G_{\vec\alpha}^F(z_{12})\ G_{\vec\alpha}^F(z_{23})\
G_{\vec\alpha}^F(z_{34})\ G_{\vec\alpha}^F(z_{45})\
G_{\vec\alpha}^F(z_{56})\ G_{\vec\alpha}^F(z_{61})\ ,\cr
g_{\vec\alpha}^{\Kc_T}(q,\vzw)&= G_{\vec\alpha}^F(z_{12})\
G_{\vec\alpha}^F(z_{23})\ G_{\vec\alpha}^F(z_{31})\
G_{\vec\alpha}^F(z_{45})\ G_{\vec\alpha}^F(z_{56})\
G_{\vec\alpha}^F(z_{64})\ ,\cr}} respectively. Thus in total, the
four different kinematics $\Kc$ receive the following one-loop
corrections: \eqn\result{ \Delta^{\rm 1-loop}_{\Kc\Tr
F^6}=\int_\Fc\fc{d^2\tau}{\tau_2^6} \fc{1}{\eta(\ov
q)^{24}}\fc{1}{\eta(q)^{12}}\ \ \sum_{\vec\alpha,\vec\be}
s_{\vec\al}\ \th_{\vec\alpha}(q)^4 \theta_{\vec\be}(\ov q)^{16}\
\int_{\Ic_\tau}\prod_{i=1}^6\  d^2 z_i \
g_{\vec\alpha}^\Kc(q,\vzw)\ f_{\vec\be}(\ov q,\vz)\ .}

\subsec{Symmetric Trace in the gauge combination}

We will first investigate one special combination of $F^6$
couplings, the symmetric trace S$\Tr F^6$. The string amplitude
\SM\ includes all permutations of gauge group generators $T^a$;
any such permutation is equal to
$\Tr(T^{a_1}T^{a_2}T^{a_3}T^{a_4}T^{a_5}T^{a_6})$ up to some
commutator terms. The commutators can be discarded if one
appropriately symmetrizes in the positions of gauge currents. Thus
extracting the (gauge) hexagonal S$\Tr F^6$ term from \SM\
amounts to averaging over 60 
permutations:\foot{For a given hexagonal diagram, which contracts the
twelve fermions in the order $(i,j,k,l,m,n)$,
there exist 5 equivalent diagrams: $(j,k,l,m,n,i),\ldots, 
(n,i,j,k,l,m)$, corresponding to the cyclic permutations. 
Furthermore, changing their orientation results in twelve equivalent
diagrams. Thus, from the $6!$ possible permutations we only take into
account $720/12=60$ hexagonal diagrams.} \eqn\extractsymm{
f_{\vec \bet}(q)={1\o 60}\hskip-3mm\sum_{(i,j,k,l,m,n)\atop 60\
{\rm permutations}} \hskip-5mm
G^F_\vbe(z_{ij})G^F_\vbe(z_{jk})G^F_\vbe(z_{kl})G^F_\vbe(z_{lm})
G^F_\vbe(z_{mn})G^F_\vbe(z_{ni}) ={-1\o 120}\ \fc{\p^6}{\p z^6}
\ln \th_\vbe(0,\tau),} with the overall group-theoretical factor
$\Tr(T^{a_1}T^{a_2}T^{a_3}T^{a_4}T^{a_5}T^{a_6})$. The last
equality is a generalization\foot{This identity can also be proven
for higher genus $\theta$-functions \STiii.} of the identity \FAY
\eqn\faybox{
B_\vbe(z_1,z_2,z_3,z_4)+B_\vbe(z_1,z_2,z_4,z_3)+B_\vbe(z_1,z_3,z_2,z_4)=
-\h\ \fc{\p^4}{\p z^4}  \ln \th_\vbe(0,\tau)\ ,} where \eqn\bfay{
B_\vbe(z_1,z_2,z_3,z_4)\equiv
G^F_\vbe(z_{12})G^F_\vbe(z_{23})G^F_\vbe(z_{34})G^F_\vbe(z_{41}).}

Thanks to the relation \extractsymm, the left-moving part
$f_\vbe(\ov q,\vz)$ of \result\ does not depend\foot{Because of
this property, it was justified to set the momenta of the
exponentials of \gauge\ to zero from the beginning. With these
exponentials no poles in momenta (in the sense of \PINCH) that
would decrease the total power in momenta are generated.} on the
vertex positions and now we may permute them also in the
functions $g_{\vec\alpha}^\Kc(q)(q,\vzw)$ without affecting the
integral. Thus we may borrow \faybox\ and \extractsymm\ to
simplify the space-time parts $g_{\vec\alpha}^\Kc(q)(q,\vzw)$ by
appropriate symmetrizations:
\eqn\spaceii{\eqalign{g_{\vec\alpha}^{\Kc_S}(q,\vzw)&=\fc{1}{6} \
\fc{\p^4}{\p z^4}  \ln \th_{\vec\al}(0,\tau)\
G^F_{\vec\alpha}(z_{56})^2 \ ,\cr
g_{\vec\alpha}^{\Kc_H}(q,\vzw)&=\fc{1}{120} \ \fc{\p^6}{\p z^6}
\ln \th_{\vec\al}(0,\tau) \ .}} For the square of the fermionic
propagator $G^F_{\vec\al}(z_{56})^2$ in
$g_{\vec\alpha}^{\Kc_S}(q,\vzw)$ we use the identity \identity.
With the same argument as outlined after Eq. \spinv, we may drop
its last term $\p^2 G_B(z_{56})$. Similar conclusions apply to
$g_{\vec\alpha}^{\Kc_L}(q,\vzw)$. Due to the symmetry property
(in the positions $z_i$) of both the space-time part and the
gauge part, the correction $\triangle_{\Kc_T}$ does not
contribute to  STr$F^6$.

Thus the one-loop corrections to the four space-time kinematics
become: \eqn\resultfinal{ \Delta^{\rm 1-loop}_{\Kc{\rm S}\Tr
F^6}=-\fc{2^6}{120}\int_\Fc d^2\tau \ \tilde
g^\Kc(q)\sum_{\vec\bet} \fc{\theta_{\vec\be}(\ov
q)^{16}}{\eta(\ov q)^{24}}\ \fc{\ov\p^6}{\p \ov
z^6}\ln\th_{\vec\bet}(0)\ ,} 
with the functions:\foot{The following
equations are obtained by using the Riemann identity (A.5), 
$\fc{\p^3}{\p z^3}\th_1(0,\tau)=-\fc{1}{32\pi^3} \eta^3E_2$, and 
$\E_2=E_2-\fc{3}{\pi\tau_2}$.}
\eqn\spacei{\eqalign{ \tilde
g^{\Kc_L}(q)&=\fc{1}{\eta^{12}}\sum_{\vec\alpha} s_{\vec\al}\
\th_{\vec\alpha}(q)^4 \lf(\fc{\p^2}{\p z^2}  \ln
\th_{\vec\al}(0,\tau)+\fc{\pi}{\tau_2}\ri)^3=\fc{(2\pi)^6}{4}\E_2\ ,\cr \tilde
g^{\Kc_S}(q)&=\fc{1}{6}\fc{1}{\eta^{12}}\sum_{\vec\alpha}
s_{\vec\al}\ \th_{\vec\alpha}(q)^4\ \fc{\p^4}{\p
z^4}\ln\th_{\vec\al}(0,\tau) \lf(\fc{\p^2}{\p z^2}  \ln
\th_{\vec\al}(0,\tau)+\fc{\pi}{\tau_2}\ri)=-\fc{(2\pi)^6}{12}\E_2\ ,\cr \tilde
g^{\Kc_H}(q)&=\fc{1}{120}\fc{1}{\eta^{12}}\sum_{\vec\alpha}
s_{\vec\al}\ \th_{\vec\alpha}(q)^4\ \fc{\p^6}{\p
z^6}\ln\th_{\vec\al}(0,\tau)=0\ ,\cr \tilde g^{\Kc_T}(q)&=0\ .}}
Note the relations: \eqn\remarkable{ L=-3S\ \ \ ,\ \ \ H=0\
,} with $L=\tilde g^{\Kc_L}(q),\ S=\tilde g^{\Kc_S}(q)$ and
$H=g^{\Kc_H}(q)$. These are exactly the same relations as they
appear at two loops \STi\ as a solution of the constraints
implied by Riemann identities. It is remarkable that they
can be derived at one loop directly.

Let us now consider the second contribution, where only eight
fermions of the gauge boson vertex operators \gauge\ are
contracted. The eight fermions stemming from those four vertex
operators give rise to the $t_8$ structure of space-time
kinematics. The other two momenta arise from the two exponentials
of the remaining two gauge vertex operators (labeled by $i$ and
$j$), contracted with their $\p X_i,\ \p X_j$: \eqn\exp{
(k_i\epsilon_j)(k_j\epsilon_i)\ \vev{\p X(z_i) X(z_j)}\vev{\p
X(z_j) X(z_i)}=-(k_i\epsilon_j)(k_j\epsilon_i)\
[\p_{z_i}G_B(z_{ij})]^2\ .} Thus these contractions will give
additional contributions to the kinematics $\Kc_L$ and $\Kc_S$,
but not to $\Kc_H$. The $t_8$ part is the same correlator that
appears in the four gauge boson amplitude. In particular, this
means that all (holomorphic) position dependence in $z_k\neq
z_i,z_j$ drops out after applying  Riemann identity on the spin
structure sum involving a product of four fermionic Green's
functions whose positions form two lines or a square. They give
the constants $\mp(2\pi )^4$, respectively. The only $z$-integral
to be done is \LSW: \eqn\tobedone{ \int_{\Ic_\tau} d^2 z_id^2
z_j\ \p_{z_i} G_B(z_i-z_j)\p_{z_j} G_B(z_j-z_i)
=\fc{4\pi^2}{3}\tau_2^2\E_2(q)\ ,} giving rise to additional
contributions to $\tilde g^{\Kc_L}$ and $\tilde g^{\Kc_S}$ of
Eq.\spacei. In order to compare these contributions with the
previous ones, we should perform the (trivial) integral over two
points $z_i$ and $z_j$ in \spacei, which gives a factor of
$(2\tau_2)^2$. Finally, we have to take into account that there
are three possibilities to obtain a given $L$ kinematics from
contracting only eight fermions. Multiplying all factors, we see
that for a given $L$ or $S$ kinematics, the contributions of
twelve-fermion contractions cancel against those of eight-fermion
contractions. To summarize, our final result is: \eqn\FINAL{
\Delta^{\rm 1-loop}_{\Kc_{L,S,H,T}{\rm S}\Tr F^6}=0\ .}

This result could have also been anticipated by observing the
following identity in the gauge sector: \eqn\gaugezero{
\sum_{\vec\bet} \theta_{\vec\be}(\ov q)^{16}\fc{\ov\p^6}{\p \ov
z^6}\ln\th_{\vec\bet}(0)=0\ .} Therefore, the vanishing of the
one-loop corrections to S$\Tr F^6$ for $SO(32)$ gauge group has
two independent explanations: one relying on the cancellations in
the gauge fermion sector and another one originating from the
cancellations in the space-time fermion sector. Of course, the
latter cancellations are more general, and allow us to generalize
our findings -- it was only the application of the symmetric
trace ``prescription'' on the gauge part, resulting in
\extractsymm, which allowed further simplifications of the
space-time part, finally resulting in the elimination of any
position dependence in the integrand \resultfinal. This procedure
does not depend on the gauge group or on group-theoretical
contractions. All our arguments from above can be applied to show
that also: \eqn\FINAL{ \Delta^{\rm 1-loop}_{\Kc_{L,S,H,T}{\rm
S}\Tr F^4\Tr F^2}=0\ \ ,\ \ \Delta^{\rm
1-loop}_{\Kc_{L,S,H,T}(\Tr F^2)^3}=0\ .}

\subsec{Generic gauge combination}

One of the main properties of the one-loop S$\Tr F^6$ coupling,
calculated in the previous subsection, was the independence of
the gauge part $f_{\vec \bet}(\ov q,\vz)$ of the vertex positions
$z_i$. This is a consequence of symmetry and protects us from
possible poles (of the kind \pinch) in the integrand, due to
massless particle exchange. Furthermore, this position
independence of the gauge part allowed substantial simplification
of the space-time part as well.

When we do not impose the symmetrized gauge trace on the gauge
part, the dependence of \hex\ on the vertex  positions does not
simplify as in Eq.\extractsymm, and the left- and right-moving
parts are coupled through the position integral \result. When one
performs these integrals explicitly, there appears one obvious
complication: the $z$-integrals under consideration contain
fermion propagators $G^F(z)$ with $z$-arguments that may take
values outside of the fundamental domain $\Ic_\tau$. The
fermionic Green's functions have the periodicity behaviour:
$G_{\vec\al}^F(z+1,q)=-e^{2\pi i\al_1}G_{\vec\al}^F(z,q)$ and
$G_{\vec\al}^F(z+\tau,q)=-e^{-2\pi i\al_2}G_{\vec\al}^F(z,q)$
under $z\ra z+1$ and $z\ra z+\tau$. Thus we pick up phases when
leaving $\Ic_\tau$. The expression for $G^F(z)$ as a power
series, used so far in the literature (see {\it e.g}. \ellisone),
\eqn\sofar{ G^F_\val(z)=2\pi i\sum_{n\in\IZ}\fc{e^{2\pi i
(n+\al_1+\h)z}}{1+(-1)^{2\al_2} \ q^{n+\al_1+\h}}  } is not
appropriate to capture these complications. It represents a
convergent power series only inside $\Ic_\tau$. These problems
can be overcome if we perform a double Fourier expansion of
\sofar. Introducing $x$ and $y$, with
$z=x+\fc{\tau_1}{\tau_2}y+iy$, {\it i.e}. $x=\re(z)-
\fc{\tau_1}{\tau_2}\im(z),\ y=\im(z)$, we perform a Fourier
expansion w.r.t.\ $x$ with period $2$ and w.r.t.\ $y$ with period
$2\tau_2$, to obtain: \eqn\fourier{
% \eqalign{
%  G^2^F(z)&=\fc{\tau_2}{2\pi^2}
%  \sum_{(m,n)}\ \fc{1}{m+\h+n\tau}\ e^{2\pi i x n}\ e^{-2\pi
%  i(m+\h)y/\tau_2} \ ,\cr
%  G^3^F(z)&=\fc{\tau_2}{2\pi^2}
%  \sum_{(m,n)}\ \fc{1}{m+(n+\h)\tau}\ e^{2\pi i x (n+\h)}\
%  e^{-2\pi i m y/\tau_2}\ ,\cr
G_{\vec\al}^F(z)= \sum_{(m,n)}\ \Lambda(s)\
\fc{1}{m+\h+\al_2+(n+\al_1+\h)\tau}\ e^{2\pi i x (n+\al_1+\h)}\
e^{2\pi i(m+\h+\al_2)\fc{y}{\tau_2}}\ .} With the regulator
\eqn\reg{ \Lambda(s)=\fc{1}{|m+\h+\al_2+(n+\al_1+\h)\tau|^{s}},}
for $s>1$, the function \fourier\ transforms manifestly
covariantly under modular transformations, and $G_{\vec\al}^F(z)$
is defined by analytic continuation to $s=0$. In this form,
$G_{\vec\al}^F(z)$ furnishes the desired properties under the
shifts $x\ra x+1$ and $y\ra y+\tau_2$ corresponding to $z\ra z+1$
and $z\ra z+\tau$, respectively.

The Fourier expansion \fourier\ is particularly convenient if the
space-time part decouples from the gauge part. In that case we
find a generating function for the integral ($d^2z=2dxdy$):
\eqn\generate{\eqalign{ \int\prod_{i=1}^N&\ d^2z_i\
G_{\vec\be}^F(z_{12})\ldots G_{\vec\be}^F(z_{N1}) =\sum_{(m,n)}\
\fc{(2\tau_2)^N}{[m+\bm+(n+\bn)\tau]^N}\cr &=2\zeta(N)(2\tau_2)^N
\lf[2^{2\bet_1N}E_N(4^{2\bet_1}\fc{\tau}{2}+\bet_1+\bet_2+\h)-E_N(\tau)\ri]\cr
&=-\fc{(2\tau_2)^N}{(N-1)!}\ \fc{\p^N}{\p
z^N}\ln\theta_{\vec\bet}(0,\tau)\ ,}} valid for even $N>2$. For
odd $N$ the integral vanishes. This relation should be compared
with the identities \faybox\ and \extractsymm.

After these preliminaries, let us first discuss the case with
only eight space-time fermions contracted. This case gives
contributions only to the kinematics $\Kc_L$ and $\Kc_S$, whose
dependence on the vertex positions is given by (cf. \exp):
\eqn\SL{ [\ \p_{z_i}G_B(z_{ij})\ ]^2\ \ G_{\vec\be}^F(\ov z_{12})\
G_{\vec\be}^F(\ov z_{23}) G_{\vec\be}^F(\ov z_{34})\
G_{\vec\be}^F(\ov z_{45})\ G_{\vec\be}^F(\ov z_{56})\
G_{\vec\be}^F(\ov z_{61}).} Here $i,j$ denote those two gauge
boson vertex operators in \SM\ whose exponentials and $\p X(z)$
contribute instead of their fermion pairs. To perform the
integral \eqn\SLint{ \Rc_{ij}\equiv -\int_{\Ic_\tau}\prod_{k=1}^6\
d^2z_k\ [\p_{z_i}G_B(z_{ij})]^2\ G_{\vec\be}^F(\ov z_{12}) \
G_{\vec\be}^F(\ov z_{23}) G_{\vec\be}^F(\ov z_{34})\
G_{\vec\be}^F(\ov z_{45})\ G_{\vec\be}^F(\ov z_{56})\
G_{\vec\be}^F(\ov z_{61})} we use the explicit expression
\fourier\ for $G_{\vec\al}^F(z)$ and the Fourier expansion for
$\p G_B(z)$, which may be found in \LSW: \eqn\dGB{ \p
G_B(z)=\sum_{(M,N)\neq\atop(0,0)}\fc{1}{M+N\tau}\fc{1}{|M+N\tau|^s}\
e^{2\pi i Nx}\ e^{2\pi i M y/\tau_2}\ .} In the following, let us
evaluate $\Rc_{ij}$ for $i<j$. The integral $\Rc_{ij}$ (performed
in the variables $x,y$) leads to various projections on the
integers of the sums $\p G_B(z_{ij})$ and $G_F(z_{rs})$. To this
end we arrive at: \eqn\TODO{\eqalign{
\Rc_{ij}&=(2\tau_2)^6\sum_{(M,N)\neq\atop(0,0)}\sum_{k,l\neq\atop(-M,-N)}
\fc{1}{M+N\tau}\ \fc{1}{M+k+(N+l)\tau}\ \fc{1}{|M+N\tau|^s}\
\fc{1}{|M+k+(N+l)\tau|^s}\cr &\times \sum_{m,n\in\IZ}
\fc{1}{[m-k+\bm+(n-l+\bn)\ov\tau]^{j-i}}\fc{1}{[m+\bm+(n+\bn)\ov\tau]^{6-j+i}}.}}
For finite $(k,l)\neq (0,0),\ (-M,-N)$ the sum over $M,N$ can be
expressed as partial fraction
$\sum\limits_{M,N\neq(0,0)}(\fc{1}{M+N\tau}-\fc{1}{M+k+(N+l)\tau})\fc{1}{k+l\tau}$,
which converges and vanishes. Therefore, non-vanishing
contributions arise only for $(k,l)=(0,0)$: \eqn\TODOi{\eqalign{
\Rc_{ij}&=(2\tau_2)^6 \sum_{(M,N)\neq\atop
(0,0)}\fc{1}{(M+N\tau)^2}\fc{1}{|M+N\tau|^s}\ \sum_{m,n\in\IZ}
\fc{1}{[m+\bm+(n+\bn)\ov\tau]^6}\cr
&=-\fc{(2\tau_2)^6}{120}\fc{\pi^2}{3}\ \E_2(\tau)\ \fc{\p^6}{\p
z^6}\ln\th_{\vec\bet}(0,\tau)\ .}} Thus, after performing the
integral \SLint, the space-time part becomes decoupled from the
gauge part. The latter is described by the second sum in \TODOi\
and may be evaluated with \generate. In that form, it becomes
obvious that it will lead to a vanishing gauge lattice sum
\gaugezero.

To obtain something potentially non-vanishing we shall take into
account all twelve fermions contracted. Similarly as in the $\Tr
F^5$ case, we first simplify the spin structure sums involving
the correlators \space\ for the four possible space-time
kinematics: 
\eqn\spinvi{\eqalign{ \tilde
g^{\Kc_L}(q,\vzw)&=\fc{1}{\eta^{12}}\sum_{\vec\al}
s_{\vec\al}\th_{\vec\al}(0)\ \lf(\fc{\p^2}{\p
z^2}\th_{\vec\al}(0,\tau)+\fc{\pi}{\tau_2}\ri)^3=\fc{(2\pi)^6}{4}\E_2\ ,\cr 
\tilde g^{\Kc_S}(q,\vzw)&=-(2\pi)^4\sum_{\vec\al} s_{\vec\al}
\fc{1}{\th_{\vec\al}(0)}
\lf(\fc{\p^2}{\p z^2}\th_{\vec\al}(0,\tau)+\fc{\pi}{\tau_2}\ri)\cr
&\hv\times\fc{\th_{\vec\al}(z_{12})}{\th_1(z_{12})}
\fc{\th_{\vec\al}(z_{23})}{\th_1(z_{23})}
\fc{\th_{\vec\al}(z_{34})}{\th_1(z_{34})}
\fc{\th_{\vec\al}(z_{41})}{\th_1(z_{41})} \ ,\cr
\tilde g^{\Kc_H}(q,\vzw)&=-(2\pi)^6\eta^6\sum_{\vec\al}
 s_{\vec\al}\fc{1}{\th_{\vec\al}(0)^2}\
 \fc{\th_{\vec\al}(z_{12})}{\th_1(z_{12})}
 \fc{\th_{\vec\al}(z_{23})}{\th_1(z_{23})}
 \fc{\th_{\vec\al}(z_{34})}{\th_1(z_{34})}\cr
 &\hv\times\fc{\th_{\vec\al}(z_{45})}{\th_1(z_{45})}
 \fc{\th_{\vec\al}(z_{56})}{\th_1(z_{56})}
 \fc{\th_{\vec\al}(z_{61})}{\th_1(z_{61})}\ ,\cr 
 \tilde g^{\Kc_T}(q,\vzw)&=(2\pi)^6\eta^6\sum_{\vec\al}
 s_{\vec\al}\fc{1}{\th_{\vec\al}(0)^2}\
 \fc{\th_{\vec\al}(z_{12})}{\th_1(z_{12})}
 \fc{\th_{\vec\al}(z_{23})}{\th_1(z_{23})}
 \fc{\th_{\vec\al}(z_{31})}{\th_1(z_{31})}\cr
 &\hv\times\fc{\th_{\vec\al}(z_{45})}{\th_1(z_{45})}
 \fc{\th_{\vec\al}(z_{56})}{\th_1(z_{56})}
 \fc{\th_{\vec\al}(z_{64})}{\th_1(z_{64})}\  .}}
Again, for the square $G_F(z)^2$ we used \identity\ and
dropped its second term. The latter gives a vanishing
contribution \vanishbos\ after partial integrations over the
positions $z_i$ due to the periodicity of both $G_B(z)$ and
$f_{\vec\bet}(\ov q,\vz)$ at the boundary of $\Ic_\tau$. In the
appendix \appAv, we simplify the sums \spinvi\ by a combined
action of Riemann and Fay trisecant identities. These
manipulations result in the following expression: \eqn\resultvvi{
\Delta^{\rm 1-loop}_{\Kc\Tr
F^6}=\int\limits_\Fc\fc{d^2\tau}{\tau_2^6}
\sum_{\vec\be}\fc{\theta_{\vec\be}(\ov q)^{16}}{\eta(\ov q)^{24}}\
\int\limits_{\Ic_\tau}\prod_{i=1}^6\ d^2 z_i\ \tilde
g^\Kc(q,\vzw)\ f_{\vec\be}(\ov q,\vz)\ ,} with the functions
$\tilde g^\Kc(q,\vzw)$ given in Eqs. (A.16), (A.17) and (A.18)
for the kinematics $\Kc_S, \Kc_H$ and $\Kc_T$, respectively. With
these expressions and noting \eqn\intvani{\eqalign{
\int_{\Ic_\tau}\ d^2z_jd^2z_k\ \p^2 G_B(z_j-z_k)\
f_{\vec\bet}(\ov q,\vz)=0\cr \int_{\Ic_\tau}\ d^2z_jd^2z_l\ \p
G_B(z_j-z_k)\ \p G_B(z_l-z_m)\ f_{\vec\bet}(\ov q,\vz)&=0\cr
\int_{\Ic_\tau}\ d^2z_jd^2z_l\ \p G_B(z_j-z_k)\ \p G_B(z_l-z_k)\
f_{\vec\bet}(\ov q,\vz)&=0\ ,}} it is straightforward to show
that the integrals \resultvvi\ for $\Kc_H$ and $\Kc_T$ vanish
after partial integrations over $\Ic_\tau$. The last two terms of
the function $\tilde g^{\Kc_S}$, shown in Eq.(A.16), do
not give zero after integrating them with the gauge part (by
applying \TODOi). However, they give a contribution which is
cancelled again by the relevant eight-fermion contraction
$\Rc_{ij}$ after taking into account the right factors, as
discussed in the previous section. Finally, the contribution from
$\tilde g^{\Kc_L}$ is cancelled against the term coming from the
eight-fermion contractions \TODOi.

We conclude: \eqn\RESULT{ \Delta^{\rm 1-loop}_{\Kc_{L,S,H,T}\Tr
F^6}=0} for a general hexagonal gauge contraction and the
space-time kinematics $\Kc_L, \Kc_S, \Kc_H$ and $\Kc_T$. As in
the previous section, the vanishing is an effect of cancellations
in the space-time fermion sector. Thus it holds for any gauge
group. All the previous steps, together with some partial
integrals of the kind \vanishbos, can be repeated to prove the
same thing for other group-theoretical contractions:
\eqn\FINALtwo{ \Delta^{\rm 1-loop}_{\Kc_{L,S,H,T}\Tr F^4\Tr
F^2}=0\ \ ,\ \ \Delta^{\rm 1-loop}_{\Kc_{L,S,H,T}(\Tr F^2)^3}=0\ .}

Finally, let us briefly comment on the one-loop corrections to CP-odd
$F^6$ couplings which appear in the discussion of anomaly
cancellation. These corrections have been calculated in \LSW\
and, except for the correction to $\Tr F^6$ which vanishes as a
result of \gaugezero, they receive non-vanishing contributions
from the boundary of the fundamental domain.

\newsec{Conclusions}
% \goodbreak

Type I - heterotic duality in $D{=}10$  predicts various relations and 
constraints
on $F^n$ couplings at different
string loop levels on both sides, as shown in the Table displayed in the
Introduction.
One of the basic predictions of this duality is the vanishing of
two-loop corrections to the heterotic $F^4$. We proved that this is indeed
the case in Section 2 by using the hyperelliptic approach to
genus two Riemann surfaces.
This result is related by supersymmetry to Adler-Bardeen
theorem for Green-Schwarz anomaly.

Furthermore, in Section 3, we showed that all heterotic $F^5$
terms vanish at one loop. In type I theory, there is a convincing
evidence \refs{\belgium,\brazil} that non-vanishing $F^5$ terms
appear already at the classical level. Formally, this corresponds
to 1.5 loops on the heterotic side, hence an order by order
comparison may not be appropriate in this case.

Similarly, all one-loop contributions to the heterotic (CP even) $F^6$ are
zero, as shown in Section 4. Matching na\"{\i}vely to the dual
side, this excludes such type I couplings at order
$e^{\Phi_I/2}$. Apart from the tree-level $\Tr F^6$, which was
the focus of \STi, the only room left for such terms in type I
theory is at order $e^{3\Phi_I/2}$. It corresponds to a
tree-level coupling on the heterotic side and probably vanishes
on similar grounds as $\Tr F^4$ does \gross. If this is indeed
the case, our results support the conjecture that any tree-level,
NBI type I $\Tr F^{2n}$ coupling appears in $D{=}10$ only at $n{-}1$ loops on
the heterotic side. Furthermore, the classical NBI action should
not receive quantum corrections, at least for $n\le 3$.

Several comments are in order here. The computations of \STi\
indicate that some $\Tr F^6$ terms are basically different from
the conventional BPS-saturated quantities, therefore they may
escape a na\"{\i}ve duality argument. It may be a general
pattern, that only certain kinematic structures, summarized in
superinvariants, are useful objects in the framework of
strong-weak coupling duality. In fact, a classification into
several superinvariants -- one class, which is sensitive only to
BPS states and receives corrections at a specific loop order, and
another class, which is sensitive to the full string spectrum and
is renormalized at various loop orders -- has been proposed for
eight-fermion terms in \AA. The heterotic tree-level coupling
$J_0=t_8t_8R^4-\fc{1}{8}\epsilon_{10} \epsilon_{10}R^4$, whose
coupling constant is proportional to $\zeta(3)$ \gross, receives
higher order corrections and is not appropriate for a duality
comparison in the above sense, in contrast to five other
superinvariants which  are related to anomaly cancellation
terms.  Two independent superinvariants have been argued to exist
for non-Abelian $\Tr F^6$ couplings in $N=4$, $D=4$ gauge
theories \BPT\ (see also \SS).  The recently
calculated tree-level $\Tr F^5$ couplings on type I side \brazil,
which are also proportional to $\zeta(3)$, are renormalized at
one-loop \MT. On the the other hand, following the Table and the
results from section 3, such couplings cannot exist on the
heterotic side. As argued before about $J_0$, $\Tr F^5$ couplings
on the type I side are not appropriate for a duality comparison.
% This restriction may be related to the fact,
% that the transcendental number $\zeta(3)$ may not follow from any
% simple non--Abelian BI generalization.
Thus the comparison order by order in coupling constant may be
justified only for a certain subclass of couplings. We plan to
carefully discuss this problem in the near future \STiii.

The vanishing of the heterotic $F^4$  at two loops, and of
$F^5$ and $F^6$ at one loop is a consequence of supersymmetry
encoded in Riemann identitities. Compactifications on tori do not
change these identities, therefore our results extend to
arbitrary gauge groups and group-theoretical contractions in any
heterotic string vacua with sixteen supercharges. In 
particular
they hold for $D=4$, $N=4$ and $D=3$, $N=8$ heterotic vacua.\foot{$F^4$ 
couplings have been discussed in the framework of heterotic - type I duality 
also in $D=8$ \refs{\BK,\paris,\LSWF}.
Our heterotic results, valid in $D=8$, could be useful in this context.} This
is in agreement with field theoretical arguments about $F^4$ couplings, which 
forbid
corrections beyond one-loop in $D=4$, $N=4$ \seiberg\ and the
absence of higher loop corrections in $D=3$, $N=8$ field theories
\ssi.
% Furthermore, the absence of one--loop corrections to $F^6$ in $D=4$,
% N=4 and $D=3$, N=8 heterotic vacua, is in lines of field theoretical
% investigations \ssii, which show that corrections to $F^6$ start at
% two--loops.

Finally, what can we say more about the two-loop heterotic versus
tree-level type I $\Tr F^6$?
% Because of the duality relation \duality\ a heterotic weak--coupling
% expansion corresponds to a type I strong--coupling expansion and it
% may not be enough, to only match $\Tr F^{2n}$ terms with the same
%dilaton dependence. It was the possible absence of higher and lower loop 
%corrections to
% the $\Tr F^{2n}$ coupling on both the heterotic and type I side,
% which justifies this comparison.
In this paper, we have essentially eliminated one possibility, that the 
mismatch
is due to some perturbative corrections complicating the comparison.
% In fact there is a close similarity between the one--loop results 
%\remarkable\ and
% our findings at two--loop \STi.
Furthermore, in $D{=}10$ there are no instanton corrections from
NS5 branes. Hence we are confident that we have a complete result,
at least on the heterotic side \STi. However, there may be a
subtlety on type I side. The Born-Infeld action describes open
strings stretched between D9-branes while the heterotic  action
considered here maps via duality onto the {\it full} type I
theory. The latter includes also non-perturbative states, not
included in the Born-Infeld action. More work is necessary in order to 
understand how do they affect the low-energy interactions.

\bigskip
\centerline{\bf Acknowledgments }\nobreak
\bigskip
We are grateful to Sonia Paban, Savdeep Sethi, and especially to Arkady 
Tseytlin
for a very useful correspondence concerning the results of \STi. 
Furthermore we thank Wolfgang Lerche for discussions.
This work is supported in part by the Deutsche Forschungsgemeinschaft (DFG),
the German-Israeli Foundation (GIF), and
the National Science Foundation under grant PHY-99-01057.

%\break
%%%%%%%%%%%%%%%%%%%%%%%%%%%%%%%%%%%%%%%%%%%%%%%%%%%%%%%
\appendix{\appA}{Tools for one-loop amplitudes}

\goodbreak
\subsec{Riemann identity}

The genus one Riemann identity \mumford\ reads:\foot{We refer the
reader to Ref. \stieberg\ for an account of one-loop $\theta$-functions.}
\eqn\riemann{\sum_{\delta}
s_{\vec\delta}\
\theta_{\vec\delta}(z_1)\
\theta_{\vec\delta}(z_2)\ \theta_{\vec\delta}(z_3)\ \theta_{\vec\delta}(z_4)=2\
\theta_1(z_1')\ \theta_1(z_2')\ \theta_1(z_3')\ \theta_1(z_4')\ ,}
with
\eqn\transone{
\!\pmatrix{z_1'\cr z_2'\cr z_3'\cr z_4'}\!=
{1\over 2}\pmatrix{ 1&1&1&1\cr 1&-1&-1&1\cr
1&-1&1&-1\cr 1&1&-1&-1}\!\!\pmatrix{z_1\cr z_2\cr
z_3\cr z_4}\ ,}
and the phases $s_{(0,0)}=1,\ s_{(0,\h)}=-1,\ s_{(\h,0)}=-1$ and
$s_{(\h,\h)}=1$.
The sum in \riemann\ runs over both even and
odd spin-structures.
When one focuses on a CP even string amplitude one would like to have
a similar formula with a sum over the even spin-structures only.
A slight modification of \riemann\ is the identity
\eqn\riemannii{
\sum_{\delta}\tilde s_{\vec\delta}\
\theta_{\vec\delta}(z_1)\
\theta_{\vec\delta}(z_2)\ \theta_{\vec\delta}(z_3)\
\theta_{\vec\delta}(z_4)=-2\ \theta_1(z_1'')\
\theta_1(z_2'')\ \theta_1(z_3'')\ \theta_1(z_4'')\ ,}
with the transformation
\eqn\transtwo{\!\pmatrix{z_1''\cr z_2''\cr z_3''\cr z_4''}\!=
{1\over 2}\pmatrix{ -1&1&1&1\cr 1&-1&1&1\cr
1&1&-1&1\cr 1&1&1&-1}\!\!\pmatrix{z_1\cr z_2\cr
z_3\cr z_4 }\ ,}
and the phases $\tilde s_{(0,0)}=1,\ \tilde s_{(0,\h)}=-1,\ \tilde 
s_{(\h,0)}=-1$ and
$\tilde s_{(\h,\h)}=-1$.
We may combine Eqs. \riemann\ and \riemannii\ to a sum over even spin 
structures only:
\eqn\evenriemann{\eqalign{
\sum_{\delta\ {\rm even}}s_{\vec\delta}\
\theta_{\vec\delta}(z_1)\
\theta_{\vec\delta}(z_2)\ \theta_{\vec\delta}(z_3)\ \theta_{\vec\delta}(z_4)
&=\theta_1(z_1')\ \theta_1(z_2')\ \theta_1(z_3')\ \theta_1(z_4')\cr
&-\theta_1(z_1'')\ \theta_1(z_2'')\ \theta_1(z_3'')\ \theta_1(z_4'')\ ,}}
with $z_i'$ and $z_i''$ given in \transone\ and \transtwo, respectively.

% \goodbreak
\subsec{Fay trisecant identity and odd $\th$-function relations}

In this subsection we derive some useful $\th$-function
relations. We start from Fay trisecant identity \FAY:
\eqn\oneloopfay{ \det_{i,j} Z_\val(x_i-y_j+D)=(-1)^{\h
n(n-1)}Z_\val(D)^{n-1}\ Z_\val(\sum_{i=1}^n x_i-\sum_{i=1}^n
y_i+D)\ } with some divisor $D=\sum_i q_i\xi_i$ of weight zero
($\sum_i q_i=0$), 
\eqn\ZZZ{ Z_\val(\sum_{i=1}^n
x_i-\sum_{i=1}^n y_i)=\th_\val(\sum_{i=1}^n x_i-\sum_{i=1}^n y_i)
\ \fc{\prod\limits_{i,j=1\atop i\neq j}^n
E(x_i,x_j)\prod\limits_{i,j=1\atop i\neq j}^n
E(y_i,y_j)}{\prod\limits_{i=1}^n E(x_i,y_i)}\ ,} 
and the (one-loop) ``prime form'': 
\eqn\primeform{
E(x,y)=\fc{\th_1(x-y,\tau)}{\th_1'(0,\tau)}\ .} 
Eq. \oneloopfay\
holds for both even and odd spin structures $\vec\al$. Here we
shall be interested in the odd case, i.e. $\vec\al=(\h,\h)$. We
seek for identities, which relate $\th$-functions with multiple
arguments, as they usually arise after applying the Riemann
identity \riemann\ to objects with fewer arguments. Since for
$D=0$ \oneloopfay\ becomes trivial ($Z_\val(0)\equiv
\th_\val(0),\ Z_1(z)\equiv Z_{(\h,\h)}(z),\ \th_1(z)\equiv
\th_{(\h,\h)}(z)$) for odd $\vec\al$, we shall first choose
$D=\xi_1-\xi_2$ and get rid of it later. For this choice we may
find an useful relation for the case $n=2$. We first multiply Eq.
\oneloopfay\ by $E(\xi_1,\xi_2)^2$. Then we differentiate the
resulting equation one times w.r.t. $\xi_1$ and take the limit
$\xi_1\ra\xi_2$: Because in this limit $\th_1(\xi_1-\xi_2)$
becomes zero, the derivative has to act on the latter. Thus we
obtain ($Z_1'(0)\equiv \th_1'(0)$): \eqn\newi{\eqalign{
&Z_1(x_1+x_2-y_1-y_2)Z_1'(0)=\cr
&-\lf.\fc{\p}{\p\xi}\lf\{\fc{\th_1(x_1-y_1+\xi)}{E(x_1,y_1)}
\fc{\th_1(x_2-y_2+\xi)}{E(x_2,y_2)}-\fc{\th_1(x_2-y_1+\xi)}{E(x_2,y_1)}
\fc{\th_1(x_1-y_2+\xi)}{E(x_1,y_2)}\ri\}\ri|_{\xi=0}\cr
&=\th_1'(0)^2[g(x_2-y_1)+g(x_1-y_2)-g(x_1-y_1)-g(x_2-y_2)]\ . }}
Here, the function $g(x-y)$ is defined by 
\eqn\ggg{
g(x-y)=\p\ln\th_1(x-y)\ .} It is related to the Green's function
$\vev{\tilde\psi(z)\tilde\psi(w)}=\p G_B(z-w)$
for the non-zero modes $\tilde \psi$ of odd fermions through the
equation  (cf. \Bosonic): \eqn\Bosonici{ g(z)=\p
G_B(z)-\fc{\pi}{\tau_2}(z-\ov z)\ .}

Similarly, we may proceed  in the case $n=3$. After multiplying \oneloopfay\
with $E(\xi_1,\xi_2)^3$, differentiating two times
w.r.t. $\xi_1$ and taking the limit $\xi_1\ra\xi_2$ we obtain:
\eqn\newii{\eqalign{
&Z_1(x_1+x_2+x_3-y_1-y_2-y_3)Z_1'(0)^2=
-\h\lf.\fc{\p^2}{\p\xi^2}\ \det_{i,j} 
\fc{\th_1(x_i-y_j+\xi)}{E(x_i,y_j)}\ri|_{\xi=0}\cr
&=-\th_1'(0)^3\lf[
           -g(x_2-y_1) g(x_1-y_2)+g(x_3-y_1) g(x_1-y_2)+g(x_1-y_1) 
g(x_2-y_2)\ri.\cr
&\hiii -g(x_3-y_1) g(x_2-y_2)-g(x_1-y_1) g(x_3-y_2)+g(x_2-y_1) g(x_3-y_2)\cr
&\hiii +g(x_2-y_1) g(x_1-y_3)-g(x_3-y_1) g(x_1-y_3)-g(x_2-y_2) g(x_1-y_3)\cr
&\hiii +g(x_3-y_2) g(x_1-y_3)-g(x_1-y_1) g(x_2-y_3)+g(x_3-y_1) g(x_2-y_3)\cr
&\hiii +g(x_1-y_2) g(x_2-y_3)-g(x_3-y_2) g(x_2-y_3)+g(x_1-y_1) g(x_3-y_3)\cr
&\hiii\lf.-g(x_2-y_1) 
g(x_3-y_3)-g(x_1-y_2)g(x_3-y_3)+g(x_2-y_2)g(x_3-y_3)\ri]\ . }}

\subsec{Inversion formula}
{}For $n=2$ and even $\vec\al$, Eq.\oneloopfay\ can be inverted:
\eqn\invertfay{\eqalign{
Z_\val & (z_1-z_2) Z_\val(z_3-z_4)=\cr
&\h Z_\val(0)[Z_\al(z_1-z_2+z_3-z_4)-Z_\val(z_1+z_2-z_3-z_4)+
Z_\val(z_1-z_2-z_3+z_4)]\ .}}
We shall use this relation in our spin structure sums 
as a preparation for applying the Riemann identity \evenriemann.

\goodbreak
\subsec{Spin-structure sums for $F^5$}

In this subsection, we perform the spin structure sum on the
space-time part.
In all our equations, whenever $g$ appears, we may simply replace it
by $\p G_B$ without introducing extra terms.
This reinstates the correct periodicity and modular behaviour.

\br
{\it Kinematics $\Kc_S$}
\br
Applying Riemann identities to the sum $\tilde g^{\Kc_S}$ of \spinv\
gives:
\eqn\spinVa{\eqalign{
-\tilde g^{\Kc_S}(q,\vzw)&=\fc{(-2\pi)^3}{\eta^3}
\fc{\p^2}{\p z^2}\lf.\sum_{\vec\al\ {\rm even}} s_{\vec\al}
\th_{\vec\al}(z)\
\fc{\th_{\vec\al}(z_{12})}{\th_1(z_{12})}\fc{\th_{\vec\al}(z_{23})}{\th_1(z_{23})}
\fc{\th_{\vec\al}(z_{31})}{\th_1(z_{31})}\ri|_{z=0}\cr
&=-\fc{8\pi^3}{\eta^3}\lf\{2\th_1(z/2)\fc{\th_1(z_1-z_2+\zh)}{\th_1(z_1-z_2)}
\fc{\th_1(z_2-z_3+\zh)}{\th_1(z_2-z_3)}
\fc{\th_1(z_3-z_1+\zh)}{\th_1(z_3-z_1)}\ri.\cr
&\hiii\lf.\lf.-2\th_1(z/2)\fc{\th_1(z_1-z_2-\zh)}{\th_1(z_1-z_2)}
\fc{\th_1(z_2-z_3-\zh)}{\th_1(z_2-z_3)}
\fc{\th_1(z_3-z_1-\zh)}{\th_1(z_3-z_1)}\ri\}\ri|_{z=0}\cr
&=(2\pi)^4\ [g(z_1-z_2)+g(z_2-z_3)+g(z_3-z_1)]\
.}}
\br
{\it Kinematics $\Kc_P$}
\br
After applying the inversion formula \invertfay\ for the two products
$\th_\al(z_{12})\th_\al(z_{34})$
and $\th_\al(z_{23})\th_\al(z_{45})$ in $\tilde g^{\Kc_P}$ of
\spinv\ we are ready to use Riemann identities for the spin structure sum:
\eqn\spinVb{\eqalign{
-\tilde g^{\Kc_P}(q,\vzw)&=(-2\pi)^5\eta^3\sum_{\vec\al\ {\rm even}}
s_{\vec\al}\fc{1}{\th_{\vec\al}(0)}\
\fc{\th_{\vec\al}(z_{12})}{\th_1(z_{12})}\fc{\th_{\vec\al}(z_{23})}{\th_1(z_{23})}
\fc{\th_{\vec\al}(z_{34})}{\th_1(z_{34})}\fc{\th_{\vec\al}(z_{45})}{\th_1(z_{45})}
\fc{\th_{\vec\al}(z_{51})}{\th_1(z_{51})}\cr
&=-\fc{4\pi^3}{\eta^3}\lf[-Z_1(z_1+z_2-z_3-z_5)-Z_1(z_1-z_2+z_3-z_5)\ri.\cr
&\hv +Z_1(z_1+z_2-z_4-z_5)-Z_1(z_1+z_3-z_4-z_5)\cr
&\hv\lf.-Z_1(z_1-z_2+z_4-z_5)-Z_1(z_1-z_3+z_4-z_5)\ri]\cr
&=(2\pi)^4\ \lf[-g(z_1-z_2)-g(z_2-z_3)-g(z_3-z_4)-g(z_4-z_5)-g(z_5-z_1)\ri]\ 
.}}
In the last equality we used \newi\ for $Z_1$'s.

\goodbreak
\subsec{Spin-structure sums for $F^6$}

Let us now proceed to the spin structure sums in the $F^6$ couplings.
Again, whenever $g$ appears, we may simply replace it 
by $\p G_B$ without introducing extra terms.
\br
{\it Kinematics $\Kc_S$}
\br
After using the identity \invertfay\ for the two products
$\th_\al(z_{12})\th_\al(z_{34})$
and $\th_\al(z_{23})\th_\al(z_{41})$ in $\tilde g^{\Kc_S}$ of
\spinvi, we apply Riemann identities for the spin structure sum:
\eqn\spinVIa{\eqalign{
-\fc{\tilde g^{\Kc_S}(q,\vzw)}{(2\pi)^4}&=
-\fc{\pi}{\tau_2}+\fc{\p^2}{\p z^2}\lf.\sum_{\vec\al\ {\rm even}}
s_{\vec\al}\fc{\th_{\vec\al}(z)}{\th_\al(0)}\
\fc{\th_{\vec\al}(z_{12})}{\th_1(z_{12})}
\fc{\th_{\vec\al}(z_{23})}{\th_1(z_{23})}
\fc{\th_{\vec\al}(z_{34})}{\th_1(z_{34})}
\fc{\th_{\vec\al}(z_{41})}{\th_1(z_{41})} \ri|_{z=0}\cr
&={1\o 4}\th_1'(0)^{-2}\lf[-Z_1(z_1+z_2-z_3-z_4)^2+Z_1(z_1-z_2+z_3-z_4)^2\ri.\cr
&\hv   \lf.  -Z_1(z_1-z_2-z_3+z_4)^2\ri]\cr
&\hiii -\h\lf[\p G_B(z_1-z_3)+\p G_B(z_2-z_4)\ri]\cr
&=  -\h\lf[-g(z_1-z_2)g(z_1-z_3)-g(z_1-z_2)g(z_2-z_3)-g(z_1-z_3)g(z_2-z_3)\ri.\cr
&\hiii +g(z_1-z_2)g(z_1-z_4)-g(z_1-z_3)g(z_1-z_4)+2g(z_2-z_3)g(z_1-z_4)\cr
&\hiii +g(z_1-z_2)g(z_2-z_4)-g(z_2-z_3)g(z_2-z_4)-g(z_1-z_4)g(z_2-z_4)\cr
&\hiii -2g(z_1-z_2)g(z_3-z_4)+g(z_1-z_3)g(z_3-z_4)-g(z_2-z_3)g(z_3-z_4)\cr
&\hiii \lf.+g(z_1 - z_4)g(z_3-z_4)-g(z_2-z_4)g(z_3 - z_4)\ri]\cr
&\hiii -\h[\p G_B(z_1-z_3)+\p G_B(z_2-z_4)]\cr
&\hiii -\h[g(z_1-z_3)^2+g(z_2-z_4)^2]\ .}}
For the last equality we made use of \newi. As it turns out in section 4.2,
the last two terms play an important r\^ole.

\br
{\it Kinematics $\Kc_H$}
\br
After applying the inversion formula \invertfay\ for the three products
$\th_\al(z_{12})\th_\al(z_{45})$, $\th_\al(z_{23})\th_\al(z_{56})$
and $\th_\al(z_{34})\th_\al(z_{61})$ in $\tilde g^{\Kc_H}$ of
\spinvi, we use Riemann identities for the sum:
\eqn\spinVIb{\eqalign{
-\tilde g^{\Kc_H}(q,\vzw)&=(2\pi)^6\eta^6\sum_{\vec\al\ {\rm even}} s_{\vec\al}
\fc{1}{\th_\al(0)^2}\
\fc{\th_{\vec\al}(z_{12})}{\th_1(z_{12})}\fc{\th_{\vec\al}(z_{23})}{\th_1(z_{23})}
\fc{\th_{\vec\al}(z_{34})}{\th_1(z_{34})}\fc{\th_{\vec\al}(z_{45})}{\th_1(z_{45})}
\fc{\th_{\vec\al}(z_{56})}{\th_1(z_{56})}\fc{\th_{\vec\al}(z_{61})}{\th_1(z_{61})}\cr
% &={1\o 4}\lf[-Z(z_1-z_2+z_3-z_4+z_5-z_6)-Z(z_1+z_2-z_3-z_4-z_5+z_6)\ri.\cr
% &\hiii\lf.+Z(z_1-z_2-z_3-z_4+z_5+z_6)\ri]+ \ldots\cr
%  &\hiii+ \ldots\cr
&=\h(2\pi)^4
\lf[-g(z_1-z_2)g(z_1-z_3)-g(z_1-z_2) g(z_2-z_3)-g(z_1-z_3) g(z_2-z_3)\ri.\cr
&\hiii-g(z_2-z_3)g(z_2-z_4)-2g(z_1-z_2)g(z_3-z_4)-g(z_2-z_3)g(z_3-z_4)\cr
&\hiii-g(z_2-z_4)g(z_3-z_4)+g(z_1-z_3)g(z_1-z_5)-g(z_1-z_3)g(z_3-z_5)\cr
&\hiii-g(z_3-z_4)g(z_3-z_5)+g(z_1-z_5)g(z_3-z_5)-2g(z_1-z_2) g(z_4-z_5)\cr
&\hiii-2 g(z_2-z_3)g(z_4-z_5)-g(z_3-z_4) g(z_4-z_5)-g(z_3-z_5) g(z_4-z_5)\cr
&\hiii+g(z_1-z_2)g(z_1-z_6)+2 g(z_2-z_3)g(z_1-z_6)+2g(z_3-z_4)g(z_1-z_6)\cr
&\hiii-g(z_1-z_5) g(z_1-z_6)+2g(z_4-z_5) g(z_1-z_6)+g(z_1-z_2)g(z_2-z_6)\cr
&\hiii+g(z_2-z_4)g(z_2-z_6)-g(z_1-z_6) g(z_2-z_6)-g(z_2-z_4)g(z_4-z_6)\cr
&\hiii-g(z_4-z_5) g(z_4-z_6)+g(z_2-z_6)g(z_4-z_6)-2 g(z_1-z_2)g(z_5-z_6)\cr
&\hiii-2g(z_2-z_3) g(z_5-z_6)-2g(z_3-z_4)g(z_5-z_6)+g(z_1-z_5)g(z_5-z_6)\cr
&\hiii\lf.-g(z_4-z_5) g(z_5-z_6)+g(z_1-z_6)g(z_5-z_6)-g(z_4-z_6) 
g(z_5-z_6)\ri]\ .}}
In the last equality, we made use of \newi\ and \newii.

\br
{\it Kinematics $\Kc_T$}
\br
We use the inversion formula \invertfay\ for the two products
$\th_\al(z_{12})\th_\al(z_{45})$ and $\th_\al(z_{23})\th_\al(z_{56})$
in $\tilde g^{\Kc_T}$ of
\spinvi\ and then apply Riemann identities for the spin structure sum:
\eqn\spinVIc{\eqalign{
\tilde g^{\Kc_T}(q,\vzw)&=(2\pi)^6\eta^6\sum_{\vec\al\ {\rm even}} s_{\vec\al}
\fc{1}{\th_\al(0)^2}\
\fc{\th_{\vec\al}(z_{12})}{\th_1(z_{12})}\fc{\th_{\vec\al}(z_{23})}{\th_1(z_{23})}
\fc{\th_{\vec\al}(z_{31})}{\th_1(z_{31})}\fc{\th_{\vec\al}(z_{45})}{\th_1(z_{45})}
\fc{\th_{\vec\al}(z_{56})}{\th_1(z_{56})}\fc{\th_{\vec\al}(z_{64})}{\th_1(z_{64})}\cr
% &=\h Z(z_1-z_2-z_3-z_4+z_5+z_6)\cr
% &\hii+{1\o 4}\lf[Z(z_1+z_2-z_4-z_5)Z(z_2+z_3-z_5-z_6) +
% Z(z_1-z_2+z_4-z_5)Z(z_2+z_3-z_5-z_6)\ri.\cr
% &\hii+Z(z_1-z_2-z_4+z_5)Z(z_2+z_3-z_5-z_6)+Z(z_1+z_2-z_4-z_5)Z(z_2-z_3+z_5-z_6)\cr
% &\hii-Z(z_1-z_2+z_4-z_5)Z(z_2-z_3+z_5-z_6)-Z(z_1-z_2-z_4+z_5)Z(z_2-z_3+z_5-z_6)\cr
% &\hii+2Z(z_1+z_4-z_5-z_6)Z(z_1-z_2-z_3+z_6)+2Z(z_1-z_3+z_5-z_6)Z(z_1-z_2-z_4+z_6)\cr
% &\hii+2Z(z_1-z_3+z_4-z_5)Z(z_2-z_3-z_4+z_6)+Z(z_1+z_2-z_4-z_5)Z(z_2-z_3-z_5+z_6)\cr
% &\hii-Z(z_1-z_2+z_4-z_5)Z(z_2-z_3-z_5+z_6)+Z(z_1-z_2-z_4+z_5)Z(z_2-z_3-z_5+z_6)\cr
% &\hii\lf.-2Z(z_1+z_2-z_3-z_6)Z(z_1-z_4-z_5+z_6)-
% 2Z(z_1+z_2-z_3-z_5)Z(z_2-z_4-z_5+z_6)\ri]\cr
&\hskip-1.7cm=\h(2\pi)^4
\lf[-g(z_1 - z_2)g(z_1 - z_3)+g(z_1 - z_2)g(z_2 - z_3)-g(z_1 -z_3)g(z_2 - 
z_3)\ri.\cr
&+g(z_2 - z_3)g(z_2 - z_4)-g(z_2 - z_3)g(z_3 - z_4)+g(z_2 -z_4)g(z_3 - z_4)\cr
& +g(z_1 - z_2)g(z_1 - z_5)-g(z_1 - z_2)g(z_2 - z_5)-g(z_2 - z_4)g(z_2 -z_5)\cr
& +g(z_1 - z_5)g(z_2 - z_5)-2g(z_1 - z_2)g(z_4 - z_5)+2 g(z_1 -z_3)g(z_4 - 
z_5)\cr
& -2g(z_2 - z_3)g(z_4 - z_5)+g(z_2 - z_4)g(z_4 - z_5)-g(z_2 - z_5)g(z_4- 
z_5)\cr
&+g(z_1 - z_3)g(z_1 - z_6)-g(z_1 - z_5)g(z_1 - z_6)-g(z_1 - z_3)g(z_3 -z_6)\cr
&-g(z_3 - z_4)g(z_3 - z_6)+ g(z_1 - z_6)g(z_3 - z_6)+2g(z_1 - z_2)g(z_4- 
z_6)\cr
&-2g(z_1 - z_3)g(z_4 - z_6)+2g(z_2 - z_3)g(z_4 - z_6)+g(z_3 - z_4)g(z_4- 
z_6)\cr
&+g(z_4 - z_5)g(z_4 - z_6)-g(z_3 - z_6)g(z_4 - z_6)-2g(z_1 - z_2)g(z_5- z_6)\cr
&+2g(z_1 - z_3)g(z_5 - z_6)-2g(z_2 - z_3)g(z_5 - z_6)+g(z_1 - z_5)g(z_5- 
z_6)\cr
&\lf.-g(z_4 - z_5)g(z_5 - z_6)-g(z_1 - z_6)g(z_5 - z_6)+g(z_4 - z_6)g(z_5 - 
z_6)\ri]\ .}}
Again, in the last equality we made use of \newi\ and \newii.

%%%%%%%%%%%%%%%%%%%%%%%%%%%%%%%%%%%%%%%%%%%%%%%%%%%%%%%
\goodbreak
\appendix{\appB}{Two-loop $\Tr F^4$ for a different gauge slice}

The path integral for a $g$-loop string amplitude contains an exponential
with a coupling $\int d^2z \chi T_F$ of the world-sheet gravitino $\chi$ to
the fermionic part of the stress tensor $T_F$.\foot{The PCO $Y$, discussed in 
Section 2, is related to the supercurrent $T_F$ by
$Y(z)\equiv e^{\phi(z)}T_F(z)$, with the background charge operator $e^\phi$.} 
Expanding the gravitino w.r.t.\ the basis $\{\chi^{(a)}=\delta^{(2)}(z-x_a)
\ ;\ a=1,\ldots,2g-2\}$
of $3/2$ differentials, and integrating over the supermoduli,
brings down $2g-2$ supercurrent operators $T_F(x_a)$ inserted at arbitrary
positions $x_a$ in the amplitude.
In other words, in this gauge, the result of integrating over the supermoduli 
is the
appearance of $2g-2$ insertions of the supercurrent $T_F(x_a)$.
The points $x_a$ are arbitrarily chosen on the Riemann surface.
Different choices are related by total derivatives w.r.t.\ to
the moduli of the Riemann surface. The final expression for the amplitude does
not depend on these points \VV,
however in practice, it is difficult to find a convenient choice.
{}Furthermore, the total derivatives encountered after changing the points 
$x_a$,
are not globally well defined in the moduli space and, if their boundary 
contributions
do not vanish, they cause problems.

In a recent beautiful series of papers \HP\ a new method for descending from 
the
supermoduli to the moduli space has been developed.
In this way, the ambiguities in choosing the gauge slice are avoided
and the invariance under changing the gauge slice becomes manifest.
Essentially, it introduces an additional coupling to the stress tensor $T(z)$,
in addition to the supercurrent $T_F(z)$ insertion.
The reinstated gauge slice-independence allows an
arbitrary choice of the insertion points $x_a$.
One particular choice, the so called split gauge (defined by the
vanishing of the fermion propagator $G_F(x_1,x_2)=0$ at these points) has
proven to be very efficient: the amplitudes become independent on $x_a$ at any
point in the moduli space.
This results in an extremely simple expression for the
heterotic two-loop cosmological constant (in $D{=}10$) including the
combined effect of supermoduli, superconformal ghost system
and background ghost charge.
In that case, the integral over moduli and supermoduli is expressed as a spin 
structure
dependent modular function \HP.

Due to the additional stress tensor insertions,
the correlators with $n$ vertex operators will now also involve couplings
of the vertex operators to $T(z)$ aside from the usual couplings to $T_F(z)$.
Two cases are possible: the split and the non-split. In the first case,
the vertex operators do not interact with $T(z)$ and $T_F(z)$, while in the 
second case
they do interact.
{}In the split case, a formula has been derived \HP\ for $n=4$: 
\eqn\dhoker{
{\cal S}_{t_8}=
\fc{1}{\chi_{10}(\Om)} \om{i}{z_1}\om{j}{z_2}\om{k}{z_3}
\om{l}{z_4} \sum_\vde \Xi^\vde_6(0,\Om)\theta_\vde(0,\Om)^3\ \p_i\p_j\p_k\p_l\
\th_\vde(0,\Om)\ ,}
which can be applied {\it e.g.} to the space-time part of a four gauge
boson amplitude \start.
Here $\Xi_\delta(\Om)$ is a complicated modular function  of weight six, 
defined in \HP.
The piece ${\cal S}_{t_8}$ accounts for eight-fermion contractions
coming from four vertex
operators in the zero-ghost picture in addition to the pieces coming from
the $T(z)$ and $T_F(z)$. Thus it gives order $\Oc(k^4)$ in momentum and
comprises\foot{However, in the description of \HP, there appears another,
non-symmetric kinematics which is believed to be cancelled by
a similar term from the non-split contribution.} the $t_8$ tensor in ten 
dimensions.
We accomplished to write \dhoker\ in a somewhat simpler way, by using
Siegel modular forms only (depending on arguments $\vec z=(z^1,z^2)$):
\eqn\muchbetter{\eqalign{
{\cal S}_{t_8}&=\fc{1}{\chi_{10}(\Om)} \om{i}{z_1}\om{j}{z_2}\om{k}{z_3}
\om{l}{z_4}\cr
&\times\p_{z^i}\p_{z^j}\p_{z^k}\p_{z^l}\ [8 E_{8,4}(\vec z,\Om)-\fc{2}{3}
E_4(\Om)E_{4,4}(\vec z,\Om)-\fc{4}{3}E_{4,2}(\vec z,\Om)^2]\lf.\ri|_{\vec 
z_i=0}}}
with
\eqn\siegelz{\eqalign{
E_{8,4}(\vec z,\Om)&=\sum_{\val\ {\rm even}}
\th_\val(0,\Om)^{12}\th_\val(\vec z/2,\Om)^4\ ,\cr
E_{4,4}(\vec z,\Om)&=
\sum_{\val\ {\rm even}}\th_\val(0,\Om)^4\th_\val(\vec z/2,\Om)^4\ ,\cr
E_{4,2}(\vec z,\Om)&=
\sum_{\val\ {\rm even}}\th_\val(0,\Om)^6\th_\val(\vec z/2,\Om)^2\ ,}}
and the Siegel forms $E_4(\Om)=\sum\limits_{\val\ {\rm 
even}}\th_\val(0,\Om)^8$ and
$\chi_{10}(\Om)=\prod\limits_{\val\ {\rm even}} \th_\val(0,\Om)^2$. The latter 
represents
the oscillator partition function.

To calculate the split contribution to the two-loop corrections to
$\Tr F^4$, one first observes that  ${\cal S}_{t_8}$ is
completely symmetric in the vertex positions $z_i$. Similarly to
Section 2, this allows to symmetrize over the positions in
the gauge part and to take the same combination of gauge
contractions as in Eq.\combination. In terms of genus two
$\theta$-functions, Eq.\Gauge\ reads \eqn\BB{
B_{\vbe}(z_1,z_2,z_4,z_3)=\fc{1}{\theta_\vbe(0,\Om)^4}
\fc{\theta_\vbe(z_1-z_2,\Om)}{E(z_1,z_2)}
\fc{\theta_\vbe(z_2-z_3,\Om)}{E(z_2,z_3)}\fc{\theta_\vbe(z_3-z_4,\Om)}{E(z_3,z_4)}
\fc{\theta_\vbe(z_4-z_1,\Om)}{E(z_4,z_1)}\ ,} with the two-loop
Szeg\"o kernel 
\eqn\twoszego{
\vev{\psi(z_1)\psi(z_2)}_\vbe=\fc{1}{\theta_\vbe(0,\Om)}
\fc{\theta_\vbe(z_1-z_2,\Om)}{E(z_1,z_2)}\ ,} 
and the prime form
\eqn\primetwo{
E(z_1,z_2)=\fc{\th_\val(z_1-z_2,\Om)}{h_\val(z_1)h_\val(z_2)}\ \ ,\ \ 
h_\val(z)=\sqrt{\p_i\th_\val(0,\Om)\ \om{i}{z}}\ \ }
for any odd spin-strucure $\val$.
Then, the analogue of
Eq. \combination\ becomes \FAY: \eqn\gaugefay{\eqalign{
&B_\vbe(z_1,z_2,z_3,z_4)+B_\vbe(z_1,z_2,z_4,z_3)+B_\vbe(z_1,z_3,z_4,z_2)\cr
&=-\h\om{i}{z_1}\om{j}{z_2}\om{k}{z_3} \om{l}{z_4}\
\p_{z^i}\p_{z^j}\p_{z^k}\p_{z^l}\ln\th_\vbe(0,\Om)\ ,}} with the
canonical one-forms $\omega_i(z),\ i=1,2$. This expression has
to be inserted into the gauge partition function $\chi_{10}^{-1}
\sum\limits_\vbe\th_\vbe(0,\Om)^{16}$. Thus, in the split case, the final 
expression
for the two-loop corrections to $t_8\Tr F^4$ becomes: 
\eqn\finalfour{\eqalign{ \triangle_{t_8\Tr F^4}^{\rm
2-loop}&=-\fc{1}{6}\int_{\Fc_2}
\fc{d^2\Om_{11}d^2\Om_{22}d^2\Om_{12}}{[\det\im(\Om)]^5}
\fc{1}{|\chi_{10}(\Om)|^2}\cr &\times\int\int
\om{i}{z_1}\om{j}{z_2}\om{k}{z_3}\om{l}{z_4}\ \wedge\ \om{\tilde
i}{\ov z_1}\om{\tilde j}{\ov z_2}\om{\tilde k}{\ov z_3}
\om{\tilde l}{\ov z_4}\cr
&\times\p_{z^i}\p_{z^j}\p_{z^k}\p_{z^l}\ [8 E_{8,4}(\vec
z,\Om)-\fc{2}{3} E_4(\Om)E_{4,4}(\vec
z,\Om)-\fc{4}{3}E_{4,2}(\vec z,\Om)^2] \lf.\ri|_{\vec z_i=0}\cr
&\times \sum_{\beta}\theta_\vbe(0,\ov \Om)^{16}\ \p_{\ov
z^{\tilde i}}\p_{\ov z^{\tilde j}} \p_{\ov z^{\tilde k}}\p_{\ov
z^{\tilde l}}\ln \theta_\vbe(0,\ov\Om)\ ,}} with the fundamental
region $\Fc_2$ of the genus two Riemann surface. The evaluation
of the $z_i$ integrals is straightforward. After taking a closer
look at the space-time part, one can derive the following
remarkable identities: \eqn\remarkid{\eqalign{ \p^4_{z^1}\ [8
E_{8,4}(\vec z,\Om)-\fc{2}{3} E_4(\Om)E_{4,4}(\vec
z,\Om)-\fc{4}{3}E_{4,2}(\vec z,\Om)^2]\lf.\ri|_{\vec z_i=0} &=0\
,\cr \p^3_{z^1}\p_{z^2}\ [8 E_{8,4}(\vec z,\Om)-\fc{2}{3}
E_4(\Om)E_{4,4}(\vec z,\Om)-\fc{4}{3}E_{4,2}(\vec
z,\Om)^2]\lf.\ri|_{\vec z_i=0} &=0\ ,\cr \p^2_{z^1}\p^2_{z^2}\ [8
E_{8,4}(\vec z,\Om)-\fc{2}{3} E_4(\Om)E_{4,4}(\vec
z,\Om)-\fc{4}{3}E_{4,2}(\vec z,\Om)^2]\lf.\ri|_{\vec z_i=0} &=0\
.}} This proves that $\triangle_{t_8\Tr F^4}^{\rm 2-loop}=0$ for
the split case. Since this result has its origin in the
cancellations in the space-time sector, one may conclude the
same for the two-loop corrections to other couplings: $(\Tr
F^2)^2, (\Tr F)^2 R^2, R^4, (R^2)^2$. However, as already mentioned,
this result is not complete because it proves the vanishing of two-loop
corrections for split contributions (w.r.t. the gauge choice of \HP) only.
In Section 2, we proved the vanishing of two-loop corrections to 
the $\Tr F^4$ coupling in
the hyperelliptic approach. Thus -- by an indirect argument -- the
non-split contributions must
vanish, too.

\listrefs
\end